\documentstyle[aps,prd,12pt,tighten,amssymb,amsmath,delarray,eqsecnum,cite,multirow,floats]{revtex}
\def\tr{{\rm Tr}}

\def\bea{\begin{eqnarray}}
\def\eea{\end{eqnarray}}
\def\ga{\gamma}
\def\ep{\epsilon}
\newcommand{\Etilde}{\Tilde{E}}
\newcommand{\Ktilde}{\Tilde{K}}

\newcommand{\Ebar}{\Bar{E}}
\newcommand{\Kbar}{\Bar{K}}
\newcommand{\Ebartilde}{\Tilde{\Bar{E}}}
\newcommand{\Kbartilde}{\Tilde{\Bar{K}}}

\newcommand{\ospk}{\mbox{$\widehat{{\mathit osp}}(2|2)_k$}}
\newcommand{\sunk}{\mbox{$\widehat{{\mathit su}}(N)_k$}}
\title{\vspace*{1cm}{\large{\bf
Disordered Dirac Fermions: the Marriage of Three Different Approaches}}}
\vspace{0.5cm}
\author{Miraculous
J. Bhaseen\footnote{\vspace*{-0.4cm}{\tt bhaseen@thphys.ox.ac.uk}},
J.-S. Caux\footnote{\vspace*{-0.4cm}{\tt caux@thphys.ox.ac.uk}},
Ian I. Kogan\footnote{\vspace*{-0.4cm}{\tt kogan@thphys.ox.ac.uk}},\\
 and
Alexei M. Tsvelik\footnote{\vspace*{-0.4cm}{\tt tsvelik@thphys.ox.ac.uk}}
\\ \vspace{0.5cm}{\small {\em Theoretical Physics (University of Oxford), \\ 1 Keble Road,
Oxford, OX1 3NP, U.K.}} }
\begin{document}
\maketitle
\vspace*{-8.5cm}
\begin{flushright}
  {\tt OUTP-00-09}\\
  {\tt cond-mat/0012240}
\end{flushright}
\vspace*{8cm}

\begin{center}
\today
\end{center}
\vspace{0.5cm}

\begin{abstract}
We compare the critical multipoint correlation functions for
two-dimensional (massless) Dirac fermions in the presence of a random
su$(N)$ (non-Abelian) gauge potential, obtained by three different
methods. We critically reexamine previous results obtained using the
replica approach and in the limit of infinite disorder strength and
compare them to new results (presented here) obtained using the
supersymmetric approach to the $N=2$ case. We demonstrate that this {\it m\'enage \`a
trois} of different approaches leads to identical results. Remarkable
relations between apparently different conformal field theories (CFTs)
are thereby obtained. We further establish a connection between the
random Dirac fermion problem and the $c=-2$ theory of dense
polymers. The presence of the $c=-2$ theory may be seen in all three
different treatments of the disorder.
\end{abstract}
\hspace{1cm}

\hspace{1.1cm}PACS: 72.15.Rh, 71.30.+h

\hspace{1.1cm}Key words: localization,  multifractality, conformal symmetry.


\newpage

\section{Introduction}
In this paper we use an exactly solvable model of disorder
to compare different theoretical approaches used to investigate the
critical behaviour of disordered (possibly electronic) systems. A 
prominent example of a disordered critical point, and a source of
many challenging problems, is that governing the plateau transitions in the integer
quantum Hall effect (IQHE) --- see for example \cite{Huckestein:Scaling}. Despite
 the numerous field theoretic
approaches to this problem based on either Pruisken's non-linear sigma model
\cite{Levine:sigmod,Pruisken:sigmod,Weiden:supersig} or the
 Chalker--Coddington network model \cite{Chalker:net} and its superspin
chain descendants
\cite{Read:Superspin,KondMar:97,Zirn:super,Zirnbauer:Continuum,Zirnbauer:Continuumcorr},
the nature of the critical point remains elusive and inaccessible by
perturbation theory \cite{Zirnbauer:integer,Bhaseen:Towards}. This is to be contrasted with the Ising
model with weak bond disorder
\cite{Dotsenko:Impurity,Dotsenko:2Disingimpure,Dotsenko:Perturbation}, for example,
where the disorder renormalizes to zero and may be treated
perturbatively. In order to gain insight into such problems it is imperative 
to fully exploit, and indeed develop, the methods available for treating disordered systems, and to obtain as many non-perturbative results as possible.
In low-dimensional systems we may be aided in this pursuit by the
availability of powerful techniques such as conformal
field theory \cite{BPZ,Francesco:CFT,Ketov:cft} and the Bethe ansatz \cite{Bethe:Ansatz,Korepin:Inverse}.

Recent studies of disordered critical points have included Dirac fermions subjected to various random potentials \cite{LudwigFSG,Mudry:Nonhermitian,Mudry:Nonhermitiancorr,Altland:Dwave}, the random XY
model \cite{Guruswamy:Glnn}, the plateau
transitions occurring in the spin quantum Hall effect and its relation
to critical percolation
\cite{Gruzberg:Exact,Cardy:Linking,Gurarie:Conf}, and the Nishimori
line in the random bond Ising model\cite{Gruzberg:Nishimori}. In all
of these examples the critical points are of a non-perturbative nature
and are described by non-trivial field
theories. 
Our knowledge of the critical behaviour in these theories
differs somewhat, and the further elucidation of their detailed
properties is a valuable enterprise.
Indeed, in problems of disorder related 
to the localization of quantum particles, one needs to depart from the
critical point in order to calculate the diffusion propagator or (for
problems with a singular density of states) the energy dependence of
the density of states; at present this program has been successfully
implemented only for the random XY model \cite{Guruswamy:Glnn}.

We turn our attention now to the problem at the heart of this paper, namely two-dimensional (Euclidean) Dirac fermions in a random non-Abelian
gauge potential. This model has the virtue of being amenable to a
variety of non-perturbative approaches and was originally introduced in Abelian
form as part of an attempt to describe the plateau transitions in the
IQHE \cite{LudwigFSG}. The non-Abelian version of the problem appeared
in a treatment of disordered d-wave superconductivity
\cite{Nersesyan:PRL72,Nersesyan:NPB438,Tsvelik:Anexactly} and has been the subject of deeper
investigation and refinement \cite{JS:hidden,Mudry:twodcft,Komudtsvelik,Caux2,Caux:disferm,Caux:exactmult,Bernard:perturbed,Ichinose:Lattice}.
Building on the foundations of these previous studies, we address this problem by means of three independent 
non-perturbative field-theoretic techniques: these are based upon the commonly
used replica \cite{Edwards:Replica} and supersymmetric
methods \cite{Efetov:super,Efetov:chaos} together with a third (model
specific) approach valid in the limit of strong disorder
\cite{Bernard:perturbed}. In our study of the four-point correlation
functions of the local density of states we are able to demonstrate that the three methods yield
coincident results. We note that whilst the reliability of the replica approach outside of perturbation theory has
been frequently and legitimately questioned --- see for example
\cite{Efetov:super,Efetov:chaos,Verbaarshot:Critique,Zirnbauer:anothercritique}
--- we use it here (in a quite straightforward manner) to reproduce the results of the
(mathematically more rigorous) supersymmetric approach. 
Although the implementation of the replica method remains a delicate
issue in general, and in many cases one must adopt various (replica) symmetry breaking
schemes --- see references \cite{Kamenev:Wigner,Kamenev:Level} for recent
examples in random matrix theory --- 
it is noteworthy that the disordered non-Abelian Dirac fermion problem renders itself
here to a textbook (replica) treatment ---  albeit with the
benefit of hindsight.

The structure of this paper is as follows: in section \ref{nonab} we
provide a brief outline of the disordered non-Abelian Dirac
fermion problem. In section \ref{localdos} we focus our attention on
the four-point correlation functions of the local density of
states (LDOS) as encoded in the so-called $Q$-matrix. This section is subdivided into three
main subsections, each of which is devoted to a different
non-perturbative field theoretical approach to the disordered Dirac
fermion problem; subsection
\ref{replicaapproach} deals with the replica approach, \ref{strongdisorder} deals with
the so-called strong disorder approach and \ref{superapproach} deals
with the supersymmetric approach. Within each of these subsections we
provide an outline of the theoretical approach and the form of the
appropriate $Q$-matrix, together with a discussion of 
the relevant conformal dimensions and four-point correlation functions.  In
order to increase the transparency of these subsections we have relegated
many of the important (but arguably involved) technical details on the
solution of the relevant WZNW models into a rather substantial appendix. The
interested reader will find gathered in this appendix the solutions to
the Knizhnik--Zamolodchikov equations which arise in this work. In
section \ref{dense} we continue our study of the supersymmetric
approach and provide a remarkably simple free field representation of
the disorder averaged theory comprising of a two-component symplectic fermion and a pair of free bosons (one compact and
the other non-compact). The two-component symplectic
fermion with central charge $c=-2$ arises in the theoretical description of dense polymers and we
demonstrate that the twist operators of the latter theory
play a fundamental r\^ole in the non-Abelian Dirac fermion problem. 
In section \ref{convergence} we discuss the convergence of approaches
to the disordered non-Abelian Dirac fermion problem.
Finally we present concluding remarks and technical appendices. In
particular, appendix \ref{sunapp} is devoted to the
$\widehat{su}(N)_{k}$ WZNW model and is subdivided into our considerations
of the $\widehat{u}(0)_k$ WZNW model --- arising in the replica
approach --- and the $\widehat{su}(N)_{-2N}$ WZNW
model --- arising in the strong disorder approach. Appendix \ref{osp22k}
is devoted to the $\widehat{osp}(2|2)_{k}$ WZNW
model which arises in the supersymmetric approach.      
\section{Non-Abelian Dirac Fermions}
\label{nonab}
We consider $N_C$ colours of two-dimensional massless Dirac fermions minimally coupled
to a non-Abelian gauge field $A_\mu\in\mathit{su}(N_C)$ : 
\begin{equation}
\label{Dirac}
S = \int d^2 \xi \sum_{\alpha,\beta=1}^{N_C}{\bar \psi}^{\alpha}
{\not \! \! D}^{\alpha \beta} \psi^{\beta},
\end{equation}
where the gauge covariant derivative takes the form
\begin{equation}
{\not \! \! D}^{\alpha \beta}=\gamma^\mu(\delta^{\alpha\beta}\partial_\mu-iA_\mu^{\alpha\beta}).
\end{equation}
The $\gamma^{\mu}$ matrices form a two-dimensional  representation of
the Clifford algebra $\{\gamma^\mu,\gamma^\nu\}=2g^{\mu\nu}$ with
Euclidean metric $g^{\mu\nu}={\rm diag}(1,1)$, and the gauge field
$A_{\mu}^{\alpha\beta}=A_{\mu}^a \tau_a^{\alpha\beta} $ may be expanded in terms of the generators $\tau_a$ of $\mathit{su}(N_C)$.
Physical quantities are obtained by disorder averaging products of
Green's functions. We use the distribution functional
\begin{equation}
\label{GaugeDist}
P[A_\mu]\propto e^{-S[A_\mu]}, \quad S[A_\mu]=\frac{1}{g_A}\int d^2\xi\,
\mathrm{Tr} \,A_\mu(\xi)A_\mu(\xi)
\end{equation}
representing the usual choice of $\delta$-correlated Gaussian white
noise for the random vector potential. In section \ref{localdos} we
shall compare three different approaches for evaluating the disorder
averaged correlation functions.

\section{Disorder Averaged Correlation Functions}
\label{localdos}
The local density of states (LDOS) of the Dirac fermion problem is
given by \cite{Nersesyan:PRL72,Nersesyan:NPB438,Tsvelik:Anexactly}
\begin{equation}
\label{ldos}
\rho({\bf r}) =
\sum_{\alpha=1}^{N_C}\left[R_{\alpha}^\dagger\,L_{\alpha}+L_{\alpha}^\dagger
R_{\alpha}
\right].
\end{equation}
In  sections \ref{replicaapproach}--\ref{superapproach} we shall investigate the disorder
averaged multipoint correlation functions of the LDOS. In particular, we shall
focus our attention on the long distance properties (obtained for example
by integrating out the massive modes) encoded in the so-called $Q$-fields:
\begin{equation}
\label{Qfield}
Q\sim\sum_{\alpha=1}^{N_C}R_{\alpha}^\dagger\,L_{\alpha}, \quad Q^\dagger\sim\sum_{\alpha=1}^{N_C}L_{\alpha}^\dagger\,R_{\alpha}
\end{equation}
We shall study the critical correlation functions of the
$Q$-fields by three different approaches; in each case our $Q$-fields
will be represented by $Q$-matrices governed by appropriate WZNW models.
\subsection{Replica Approach}
\label{replicaapproach}
In order to average over disorder we introduce $N_F$ flavours (or
replicas) of the massless Dirac fermions appearing in equation
(\ref{Dirac}):
\begin{equation}
\label{replicated}
S^{{\rm Replica}} = \int d^2 \xi
\sum_{i=1}^{N_F}\sum_{\alpha,\beta=1}^{N_C}{\bar \psi}^{\alpha, i}
{\not \! \! D}^{\alpha \beta} \psi^{\beta, i}
\end{equation}
The (replicated) action (\ref{replicated}) enjoys  a global
$\mathit{SU}(N_F)\times\mathit{SU}(N_C)\times U(1)$ symmetry and may be recast
using Witten's non-Abelian bosonization rules \cite{Witten:nonab} ---
for reviews of the bosonization approach see for example \cite{FrishSonnen:Boson,Tsvelik:boson}. It is well known that
the free Dirac action (in the absence any gauge fields) with the flavour-colour symmetry described above
may be represented as the sum of three critical
Wess--Zumino--Novikov--Witten (WZNW) models of the form 
 \begin{equation}
\label{WZNW}
W_k[g]=\frac{k}{16\pi}\int d^2\xi\, \tr^\prime(\partial^\mu g^{-1}\partial_\mu
g)+k \Gamma[g]
\end{equation}
where (the WZNW term) $\Gamma$ reads
\begin{equation}
\label{WZNWterm}
\Gamma[g]=\frac{i}{24\pi}\int d^3x \,\epsilon^{\mu\nu\rho}\,\tr^\prime(g^{-1}\partial_\mu g\, g^{-1}\partial_\nu g\, g^{-1}\partial_{\rho}g). 
\end{equation}
The celebrated equivalence between the free (flavoured-coloured) Dirac fermion theory and
the bosonic WZNW models may be written in the symbolic
form
\begin{equation}
\mbox{Free Dirac Fermions} = 
\widehat{\mathit{su}}(N_C)_{N_F}\times
\widehat{\mathit{su}}(N_F)_{N_C}\times \widehat{u}(1) \label{embed}
\end{equation}
 where it is understood that the {\it chiral} blocks of the free Dirac theory are
obtained as products of the {\it chiral} blocks of the WZNW models;
here $\widehat{g}_k$  denotes the group manifold $g$ and the so-called
level $k$ of the WZNW action (\ref{WZNW}). That this is indeed an equality
between {\it chiral} blocks is seen most clearly in the work of
Fuchs \cite{Fuchs:Free}. Of paramount importance in the
application of  (\ref{embed}) to our disordered problem
 is that the disorder --- the su$(N_C)$ gauge potential --- 
couples {\em only} to (currents from) the $\widehat{\mathit{su}}(N_C)_{N_F}$
sector. Averaging over disorder (equivalently integrating over the su$(N_C)$
gauge potential) generates a (quadratic) interaction between su$(N_C)$
currents only. This interaction scales to the strong coupling regime
where a mass gap $M\sim\exp[-2\pi/N_C\lambda]$ is dynamically
generated in the $\widehat{\mathit{su}}(N_C)_{N_F}$ sector of the theory. As follows
from the decomposition (\ref{embed}) the massless degrees of freedom are
SU$(N_C)$ singlets and are described by the remaining WZNW models
$\widehat{\mathit{su}}(N_F)_{N_C}\times u(1)\sim \widehat{\mathit{u}}(N_F)_{N_C}$. Upon integrating out the massive (colour) degrees of freedom, the fermion bilinears are expressed in terms of
the so-called $Q$-matrix
\begin{equation}
\label{repqmatrix}
Q_{r {\bar r}}\sim\frac{1}{M}\sum_{\alpha=1}^{N_C}
R^\dagger_{\alpha,r}L_{\alpha,{\bar r}}
\end{equation}
whose indices reside in the flavour (replica) space; this is simply a
replicated version of the $Q$-field introduced to describe the LDOS in
equations (\ref{ldos}) and (\ref{Qfield}). The $Q$-matrix assumes values in the group ${\rm
U}$($N_F$)
and is governed by a (critical) effective action of the WZNW form (\ref{WZNW})
with level $k=N_C$. The WZNW model thus plays an analogous r\^ole to the sigma model in the
conventional theory of localization \cite{Efetov:super} --- it is an
effective action for the slow degrees of freedom, once the fast
degrees of freedom have been integrated out.

\subsubsection{Conformal Dimensions}
In the replica approach, the conformal dimension of the $Q$-matrix is
that of a primary field of the ($N_F\rightarrow 0$) $\widehat{u}(N_F)_{N_C}$
WZNW model transforming in the fundamental representation ---
see equation (\ref{repdim}) of appendix \ref{sunapp}:
\begin{equation}
\label{ldosdim}
h_Q=\frac{1}{2N_C^2}
\end{equation}
As we shall see in subsections \ref{strongdim} and \ref{superdim}, the
conformal dimension of the $Q$-matrix (LDOS) is given by (\ref{ldosdim})
in all three treatments of the disorder. This ensures the equality of
their two-point and three-point correlation functions. It is
well known in CFT however, that the four-point correlation functions
are {\it not}
determined solely by scaling dimensions, but in general have a
non-trivial dependence on the so-called anharmonic ratios
(\ref{anharm}) \cite{Polyakov:Conf}; in CFTs of the WZNW form,
this dependence is obtained by solving the appropriate
Knizhnik--Zamolodchikov equations \cite{Knizam:current}. In order to
compare different theoretical approaches to the random Dirac fermion problem it
is thus {\it essential} to study the
four-point correlation functions of the $Q$-matrix.
 
\subsubsection{Correlation Functions of the $Q$-field}
The correlation functions of the $Q$-matrix are those pertaining to the
${\widehat u}(N_F)_{N_C}$ WZNW model in the replica ($N_F\rightarrow
0$) limit, and are thus obtained by solving the ${\widehat u}(N_F)_{N_C}$
Knizhnik--Zamolodchikov equations \cite{Knizam:current}. This fact was
recognised in the early replica treatment of the random Dirac fermion problem
\cite{Nersesyan:PRL72,Nersesyan:NPB438}. The approach  was
subsequently refined to accommodate the  
logarithmic solutions to the Knizhnik--Zamolodchikov equations which
appear in the replica
limit \cite{JS:hidden}. Despite the advances made in
these works they suffer (in places) from a na{\"\i}ve implementation of the
replica trick; an extraneous trace over replica indices
was performed in a number of instances. In order to address these issues, and
to demonstrate the
convergence of approaches to the random
Dirac fermion problem, we rederive a number of results and adopt a
uniform notation throughout. 

The four-point correlation function of
the $Q$-matrix admits the $U(N_F)\times U(N_F)$ invariant
decomposition \cite{Knizam:current,JS:hidden} (see appendix
\ref{unkwznw}):
\begin{equation}
\label{repfpt}
\langle Q_{r_1{\bar r}_1}(1) Q_{r_2{\bar r}_2}^\dagger(2) Q_{r_3{\bar
r}_3}^\dagger(3) Q_{r_4 {\bar
r}_4}(4)\rangle=|z_{14}z_{23}|^{-2/N_C^2}\sum_{i,j=1}^2I_i^{r}{\bar
I}_j^{{\bar r}}F_{ij}(z,\bar z)
\end{equation}
where $r$  denotes the ordered sequence
of flavour (replica) indices $r_1,r_2,r_3,r_4$, and where the invariant
tensors $I_1$ and $I_2$ are defined as
$I_1^{r}=\delta^{r_1,r_2}\delta^{r_3,r_4}$ and
$I_2^{r}=\delta^{r_1,r_3}\delta^{r_2,r_4}$, together with similar equations for  ${\bar I}_1$ and ${\bar I}_2$.
The
anharmonic ratio $z$ is defined as $z=z_{12}z_{34}/z_{14}z_{32}$ and
similarly for $\bar z$.
The functions $F_{ij}(z,{\bar z})$ are single-valued combinations of
the solutions to the $\widehat{u}(N_F)_{N_C}$ Knizhnik--Zamolodchikov equations, which in the
replica limit ($N_F\rightarrow 0$) are given by equations
(\ref{replicafij}) with $k=N_C$. 
For example, setting all replica indices equal to $1$ yields
\begin{gather}
\label{qdiag}
\langle Q_{1\bar 1}(1) Q_{1\bar 1}^\dagger(2) Q_{1\bar
1}^\dagger(3) Q_{1\bar 1}(4)\rangle \sim |\Upsilon|^2\left[K_{N_C}(z)K_{N_C}(1-\bar
z)+K_{N_C}(1-z)K_{N_C}(\bar z)\right]
\end{gather}
where $\Upsilon=[z_{14}z_{23}\,z(1-z)]^{-1/N_C^2}$ and $K_{N_C}(z)$ and $E_{N_C}(z)$ are natural generalizations of the
complete elliptic integrals --- see equation (\ref{genelliptic}). In
addition one may consider `mixed' correlation functions of the form:
\begin{gather}
\label{qoffdiag}
\langle Q_{1\bar 1}(1) Q_{1\bar 1}^\dagger(2) Q_{2\bar
2}^\dagger(3) Q_{2\bar 2}(4)\rangle \sim |\Upsilon|^2\left[E_{N_C}(z)(K_{N_C}(1-\bar
z)-E_{N_C}(1-\bar z))+ c.c.\right]
\end{gather}
where $c.c$ stands for complex conjugation --- replacement of $z$ by
$\bar z$. Correlation functions calculated with respect to this
effective (replicated) action correspond to correlation functions
averaged with respect to the initial action with quenched disorder
\cite{Bernard:perturbed,Davis:random}:
\begin{subequations}
\label{repinterpret}
\begin{eqnarray}
\overline{\langle Q(1) Q^\dagger(2) Q^\dagger(3) Q(4) \rangle}_{A} &
= &
\langle Q_{r_i\bar{r_i}}(1) Q_{r_i\bar{r_i}}^\dagger(2) Q_{r_i\bar{r_i}}^\dagger(3)
Q_{r_i\bar{r_i}}(4)\rangle_{\mathrm{rep}} \label{samerep}\\
\overline{\langle Q(1) Q^\dagger(2) \rangle_A\langle Q^\dagger(3) Q(4)
\rangle}_{A} & = &
\langle Q_{r_i\bar{r_i}}(1) Q_{r_i\bar{r_i}}^\dagger(2) Q_{r_j\bar{r_j}}^\dagger(3)
Q_{r_j\bar{r_j}}(4)\rangle_{\mathrm{rep}}, \quad r_i\neq r_j \label{mixedrep}
\end{eqnarray}
\end{subequations}
The left hand side of (\ref{samerep}) represents the four-point
correlation function of the $Q$-field (calculated with respect to the original
action in the presence of quenched disorder) averaged over disorder (denoted by an
overline); the left hand side of (\ref{mixedrep}) represents the
product of two two-point
correlation function of the $Q$-field (again calculated with respect to the original
action in the presence of quenched disorder) averaged over disorder.

We note that as a direct consequence of the rather simple replica index structure of
(\ref{repfpt}) the results (\ref{samerep}) and (\ref{mixedrep}) are independent of which
replicas are actually considered; this is a manifestation of replica
symmetry --- see for example \S 3.3 of reference \cite{Fisher:Spin}.
The correlation functions of the LDOS may be
obtained from these results by means of the decomposition
(\ref{ldos}) together with crossing symmetry. We emphasize that one does {\it not} perform a trace over
replica indices in order to extract the LDOS; the traces over replica indices appearing in
equation (60) of reference \cite{Nersesyan:NPB438} and equation (5) of
reference \cite{JS:hidden} are erroneous. We further draw attention to the
simplicity and manifest crossing symmetry of the results (\ref{qdiag}) and (\ref{qoffdiag}) as
compared with those obtained in reference \cite{JS:hidden}.

\subsection{Strong Disorder Approach}
\label{strongdisorder}
As was first discussed in \cite{Bernard:perturbed,Mudry:twodcft}, and
subsequently developed in \cite{Caux:disferm}, the random Dirac
fermion problem is amenable to a direct treatment in the limit of
infinite disorder strength ($g_A\rightarrow\infty$) without invoking
either replicas or
supersymmetry. We note that in this limit the probability measure ({\ref{GaugeDist}}) is
absent. We summarize here only the important details. We separate the action (\ref{Dirac}) into its chiral components
\begin{equation}
S=\int d^2\xi
\sum_{\alpha=1}^{N_C}\left[R^\dagger_\alpha(\bar\partial+i\bar A)R_{\alpha}+L^\dagger_\alpha(\partial+iA)L_{\alpha}\right]
\end{equation}
and parameterize the gauge fields in terms of group elements $g(\xi)\in
SU^{\Bbb C}(N)\sim SL(N;{\Bbb C})$
residing in the complex extension of $SU(N)$  
\begin{equation}
\label{gaugeparam}
A=i\partial g g^{-1}, \quad \bar A=i\bar\partial {\bar g}{\bar g}^{-1}.
\end{equation}
This enables one to perform the chiral gauge transformations
\begin{equation}
\label{chiralgts}
 L\rightarrow g{\mathcal L}, \quad R\rightarrow {\bar g}{\mathcal R},
\end{equation}
so as to render a theory of free fermions (${\mathcal R},{\mathcal L}$)
decoupled from the gauge fields. An important feature of this procedure is
that the Jacobian of the transformations (\ref{chiralgts}) is proportional to the partition
function of the original action (\ref{Dirac}) at fixed disorder,
$Z[A_\mu]$ ---  this cancels the normalizing partition function
in (fixed disorder) correlation functions and removes the need to invoke
replicas or supersymmetry in order to perform disorder averaging. 
The Jacobian associated with the change of variables (\ref{gaugeparam}) is well known to
involve the WZNW action (\ref{WZNW}) on the (non-compact) manifold $h=g^\dagger g\in
SU^{\Bbb C}(N)/SU(N)$ at level $k=-2N_C$ \cite{Poly:nonab,Gawedzki:coset,Bernard:perturbed,Mudry:twodcft}:
\begin{equation}
{\mathcal D}A={\mathcal D}g \,\exp\left({2N_CW_{-2N_C}[g^\dagger g]}\right)
\end{equation}
In reference \cite{Caux:disferm} the Wakimoto free-field representation of the
$SU^{\Bbb C}(N)/SU(N)$ WZNW model at $k=-2N$ was constructed, and it was shown to
generate the $\widehat{{\mathit su}}(N)_{-2N}$ Ka{\v c}--Moody algebra;
in other words, the correlation functions of the $h$-fields may be
obtained from the solution of the $\widehat{{\mathit
su}}(N)_{-2N}$ Knizhnik--Zamolodchikov equations --- see appendix
\ref{sunminus2n}. In this approach the $Q$-fields of the random Dirac
fermion problem with $N_C$ colours are expressed as primary
fields of the 
$\widehat{{\mathit su}}(N_C)_{-2N_C}$ WZNW model `dressed' by free fermions:
\begin{equation}
\label{qstrong}
Q=\sum_{\alpha,{\bar\alpha}=1}^{N_C}{\mathcal R}_\alpha^\dagger\,
h_{\alpha{\bar\alpha}}\,{\mathcal L}_{{\bar\alpha}}, \quad
Q^\dagger=\sum_{\alpha,{\bar\alpha}=1}^{N_C}{\mathcal L}_{\bar \alpha}^\dagger\,
h_{{\bar \alpha},\alpha}^\dagger \,{\mathcal R}_{\alpha},
\end{equation}
We shall study the conformal dimensions and correlation functions of
these fields in the subsections below.
\subsubsection{Conformal Dimensions}
\label{strongdim}
It is readily seen from the strong disorder decomposition (\ref{qstrong}) that the conformal dimension of
the $Q$-field is that of free Dirac fermion ($h=1/2$) and an
$\widehat{{\mathit su}}(N_C)_{-2N_C}$ primary field --- see equation
(\ref{dimensionsu}) and set $N={N_C}$ and $k=-2N_C$:
\begin{equation}
\label{qstrongdim}
h_Q=\frac{1}{2}+\frac{N_C^2-1}{2N_C(N_C-2N_C)}=\frac{1}{2N_C^2}
\end{equation}
This agrees with the replica result (\ref{ldosdim}) and ensures the
equality of the two-point and three-point correlation functions in
both approaches. The coincidence of higher correlation functions will
be discussed below.

\subsubsection{Correlation functions of the $Q$-field}
The correlation functions of the $Q$-field are thus obtained by a
fermionic `dressing' of the correlation functions of the
$\widehat{{\mathit su}}(N_C)_{-2N_C}$ WZNW model. We note that whilst the chiral solutions to the
$\widehat{{\mathit su}}(N_C)_{-2N_C}$ WZNW model given in
\cite{Caux:disferm} are correct, their normalization constants are erroneous --- they do not satisfy
the conformal bootstrap. We provide the correct conformal blocks in
equations (\ref{strongblocks}) and their expression in terms of
generalized elliptic integrals in equations (\ref{strongblocksellip}). Moreover, in the light of the new results
obtained in the replica
approach, we shall find it convenient to generalize the decomposition
(\ref{qstrong}) slightly so as to accommodate disorder averages of {\it
products} of quenched correlation functions such as those appearing in
equation (\ref{mixedrep}). To this end we introduce as many {\it
additional} fermionic species (denoted by an index $p$) as quenched correlation
functions we wish to disorder average. We emphasize that these additional
indices are {\it not} required to {\it perform} the disorder
averaging (as would be true of replicas) but simply encode which combinations of  quenched correlation functions are to
be averaged over disorder --- in this approach we do {\it not} perform an $N_p\rightarrow
0$ limit. The $Q$-matrix acquires a pseudo-replica
index structure (labelled by the index $p$) analogous to that appearing in equation (\ref{repqmatrix}):
\begin{equation}
\label{qstrongrep}
Q_{p,\bar p}=\sum_{\alpha,{\bar\alpha}=1}^{N_C}{\mathcal R}_{\alpha,p}^\dagger\,
h_{\alpha,{\bar\alpha}}\,{\mathcal L}_{{\bar\alpha},{\bar p}}, \quad
Q_{p,\bar p}^\dagger=\sum_{\alpha,{\bar\alpha}=1}^{N_C}{\mathcal
L}_{\bar \alpha,{\bar p}}^\dagger\,
h_{{\bar \alpha},\alpha}^\dagger \,{\mathcal R}_{\alpha, p},
\end{equation}
The correlation functions of the fields (\ref{qstrongrep}) may be
written in the form
\begin{equation}
\label{strongqcorr}
 \langle Q_{p_1{\bar p}_1}(1) Q_{p_2{\bar p}_2}^\dagger(2) Q_{p_3{\bar
p}_3}^\dagger(3) Q_{p_4 {\bar
p}_4}(4)\rangle  = 
 |z_{14}z_{23}|^{-4h}\sum_{a,b=1}^2
C_{ab}\,{\mathcal G}_{p}^{(a)}(z_i)\,{\mathcal
G}_{{\bar p}}^{(b)}({\bar z}_i)
\end{equation}
where $h=(1-N_C^2)/2N_C^2$ is the conformal dimension of the
$\widehat{{\mathit su}}(N_C)_{-2N_C}$ primary field
$h_{\alpha,\bar\alpha}$ as given by equation (\ref{dimensionsu}). ${\mathcal
G}_{p}^{(a)}(z_i)$ is the (holomorphic) product of the free Dirac and $\widehat{{\mathit
su}}(N_C)_{-2N_C}$ correlation functions traced over the colour indices:
\begin{equation}
\label{strongproducts}
{\mathcal G}_{p}^{(a)}(z_i)=\sum_{\alpha=1}^{N_C}\left(\frac{I_1^\alpha I_1^p}{z_{12}z_{34}}-\frac{I_2^\alpha
I_2^p}{z_{13}z_{24}}\right)\times\left(I_1^{\alpha}F_1^{(a)}(z)+I_2^{\alpha}F_2^{(a)}(z)
\right)
\end{equation}
The coefficients $C_{ab}$ appearing in equation (\ref{strongqcorr}) have the
values $C_{12}=C_{21}=1$, $C_{11}=C_{22}=0$ to ensure
single-valuedness and their off-diagonal
form reflects the logarithmic nature of the underlying $\widehat{{\mathit
su}}(N_C)_{-2N_C}$ LCFT. We use the symbol $p$  here to denote the ordered sequence
of pseudo-replica indices $p_1,p_2,p_3,p_4$ and the invariant tensors
$I_1$ and $I_2$ are defined as
$I_1^p=\delta^{p_1,p_2}\delta^{p_3,p_4}$,
$I_2^p=\delta^{p_1,p_3}\delta^{p_2,p_4}$; analogous expressions hold for $\alpha$. 
Performing the trace over colour indices appearing in (\ref{strongproducts}) by means of the identities
\begin{equation}
\sum_{\alpha=1}^{N_C}I_{1}^{\alpha}I_1^\alpha=N_C^2 \quad \sum_{\alpha=1}^{N_C}I_{1}^{\alpha}I_2^\alpha=N_C
\end{equation}
and utilizing the relations (\ref{strongtraceblocks}) satisfied by the $\widehat{{\mathit
su}}(N_C)_{-2N_C}$ chiral blocks one obtains:
\begin{subequations}
\label{gellip}
\begin{align}
{\mathcal G}_{p}^{(1)}(z_i) & =  \frac{c^\prime}{z_{14}z_{32}}\,\Lambda\bigl\{I_1^pE_{N_C}(z)+I_2^p\bigl[K_{N_C}(z)-E_{N_C}(z)\bigr]\bigr\}\\
{\mathcal G}_{p}^{(2)}(z_i) &  =  \frac{2c^\prime N_C}{z_{14}z_{32}}\,\Lambda\left\{I_1^p\left[{\Tilde E}_{N_C}(z)-{\Tilde K}_{N_C}(z)\right]-I_2^p
{\Tilde E}_{N_C}(z)\right\}
\end{align}
\end{subequations}
where $c^\prime=cN(N^2-1)$ is a normalization constant, $\Lambda=[z(1-z)]^{-1/N_C^2}$, and where we have adopted the notation that  ${\Tilde f}(z)\equiv f(1-z)$ for an arbitrary
function $f(z)$. In particular by  inserting the explicit results
(\ref{gellip}) into the decomposition (\ref{strongqcorr}) and
collecting the coefficients of $I_i^p{\bar I}_j^{\bar p}$ one may recast
(\ref{strongqcorr}) so as to read:
\begin{equation}
\label{strongrecast}
\langle Q_{p_1{\bar p}_1}(1) Q_{p_2{\bar p}_2}^\dagger(2) Q_{p_3{\bar
p}_3}^\dagger(3) Q_{p_4 {\bar
p}_4}(4)\rangle=-\frac{2{c^\prime}^2}{X_{12}}|z_{14}z_{23}|^{-2/N_C^2}\sum_{i,j=1}^2I_i^{p}{\bar
I}_j^{{\bar p}}F_{ij}(z,\bar z)
\end{equation}
where the $F_{ij}(z,\bar z)$ are (both fortunately and remarkably) single-valued combinations of
the solutions to the $\widehat{u}(0)_{N_C}$ Knizhnik--Zamolodchikov
equations as  given by equations
(\ref{replicafij}) with $k=N_C$. That is to say (upto an irrelevant
normalization) we have recovered the replica result (\ref{repfpt}) in which
our psuedo-replica indices play the r{\^o}le of replica indices. In
particular the results (\ref{qdiag}) and (\ref{qoffdiag}) together
with their interpretations (\ref{repinterpret}) follow straightforwardly.

\subsection{Supersymmetric Approach}
\label{superapproach}
In the supersymmetric approach to disordered fermionic systems one
introduces bosonic copies of the original Grassmann fields
\cite{Efetov:super,Efetov:chaos}. The partition function of the
resulting supersymmetric theory is
equal to unity due to the inverse relationship between ordinary $c$-number
Gaussian functional integrals and their Grassmann counterparts. The
absence of a disorder dependent partition function normalizing quenched correlation functions drastically
simplifies their disorder averaging. The supersymmetric approach to the
random Dirac fermion problem has been outlined by Bernard and LeClair
\cite{Bernard:spincharge} who, following the general principles of the
supersymmetric approach, have introduced bosonic
copies of the Grassmann fields coupled to the same gauge
potential. The resulting action is given by (\ref{replicated}) where the summation over replicas is
replaced by a summation over fermionic (Grassmann) and bosonic ($c$-number) fields:
\begin{equation}
\label{supersymmetrized}
S^{{\rm SUSY}} = \int d^2 \xi
\sum_{i=1}^{2}\sum_{\alpha,\beta=1}^{N_C}{\bar \psi}^{\alpha, i}
{\not \! \! D}^{\alpha \beta} \psi^{\beta, i}
\end{equation}
The symmetry of the free supersymmetrized Dirac action (in the absence of
any gauge fields) is $OSp(2N_C|2N_C)$ \cite{Bernard:perturbed} and it
may be recast as the ${\widehat{osp}}(2N_C|2N_C)_1$ WZNW model. In
particular, the
random $su(N_C)$ gauge potential couples only to currents $J^a$ (and
${\bar J}^a$) residing in
the ${\widehat{su}}(N_C)_0$ subalgebra of the complete
${\widehat{osp}}(2N_C|2N_C)_1$ Ka{\v c}--Moody algebra:
\begin{equation}
\label{supersym}
S^{{\rm SUSY}}=\widehat{osp}(2N_C|2N_C)_1+\int d^2\xi \,(J^a A^a+{\bar J}^a {\bar A}^a)
\end{equation}
In the special case $N_C=2$, Bernard and LeClair have demonstrated that the
Suguwara energy momentum tensor for ${\widehat{osp}}(2N_C|2N_C)_1$ may
 be decomposed into the sum of two commuting pieces pertaining to
different symmetries \cite{Bernard:spincharge}:
\begin{equation}
\label{sugsep}
T_{\rm{osp}(4|4)_1}=T_{\rm{osp}(2|2)_{-2}}+T_{{\rm su}(2)_0}
\end{equation}
Exploiting the decomposition (\ref{sugsep}) one may rewrite equation
(\ref{supersym}) in the {\it decoupled} form
\begin{equation}
\label{susydecoup}
S^{{\rm SUSY}}= \widehat{osp}(2|2)_{-2}+\widehat{su}(2)_{0}+ \int d^2\xi\, (J^aA^a+{\bar J}^a{\bar A}^a)
\end{equation}
We note that the r{\^o}le of the decomposition (\ref{sugsep}) in this
approach closely
mirrors that  of the decomposition (\ref{embed}) employed in the replica
approach ---  both allow us to decouple the effects of the gauge potential disorder. Indeed, the latter two terms appearing in equation (\ref{susydecoup})
are precisely those which appear in the replica approach as
$N_F\rightarrow 0$.  As was
rigorously proven in references
\cite{Nersesyan:NPB438,Nersesyan:PRL72}, and utilized to our advantage
in section
\ref{replicaapproach}, this $su(2)$ sector becomes
massive for the simple Gaussian distribution of the gauge fields given
in equation (\ref{GaugeDist}); a perturbative renormalization group
argument was given in reference \cite{Bernard:spincharge} where the one loop beta
function was calculated with the result that the coupling constant $g_A$ flows to
strong coupling.\footnote{Although a formal proof of the gap
generation exists only for the case when the disorder distribution is
Gaussian, it appears reasonable to assume that the $su(2)$ sector
remains massive for a broader choice of disorder distributions. At
scales smaller than the gap, the effective action is given by the
critical ${\widehat{osp}}(2|2)_{-2}$ WZNW model. The critical
point is stable with respect to variations of the disorder: all such
variations affect only the massive modes and hence do not generate
relevant perturbations of the critical action.} That is to say, the
low-energy effective theory governing the $Q$-field is the
${\widehat{osp}}(2|2)_{-2}$ WZNW model. Fortunately, the ${\widehat{osp}}(2|2)_{k}$ WZNW model has been discussed quite
extensively in the work of Maassarani and Serban
\cite{Maassarani:non}. In particular the model
undergoes a dramatic simplification at $k=-2$ \cite{Bhaseen:legendre}
and we provide a rather
extensive discussion of this model in Appendix \ref{osp22k}. In this
approach the $Q$-field may be represented as
\begin{equation}
\label{Qsuper}
Q=Q^{1,{\bar 1}}, \quad Q^\dagger=Q^{4,{\bar 4}}
\end{equation}
where the $Q^{\alpha,\bar\alpha}$ are primary fields transforming in the
$[0,1/2]$ representaion of $osp(2|2)$ --- see Appendix \ref{osp22k}.
\subsubsection{Conformal Dimensions}
\label{superdim}
In the supersymmetric approach to the random $su(2)$ Dirac fermion
problem, the conformal dimension of the
$Q$-field coincides with that of a primary field transforming in the
fundamental $[0,1/2]$ representation of
${\widehat{osp}}(2|2)_{-2}$ --- see equation (\ref{ospdim}) and set $k=-2$:
\begin{equation}
h_Q=\frac{1}{4-2(-2)}=\frac{1}{8}
\end{equation}
This is in agreement with the replica result (\ref{ldosdim}) and the strong
disorder result (\ref{qstrongdim}) when $N_C=2$.

\subsubsection{Correlation functions of the $Q$-field}
The four-point correlation function of
the $Q$-matrix admits the $osp(2|2)\times osp(2|2)$ invariant
decomposition --- see equations (\ref{superfour}), (\ref{globalfourpt}) and (\ref{fij}):
\begin{equation}
\langle Q^{\alpha_1,{\bar \alpha}_1}(1) Q^{\alpha_2,{\bar
\alpha}_2}(2) Q^{\alpha_3,{\bar \alpha}_3}(3) Q^{\alpha_4,{\bar
\alpha}_4}(4)\rangle=|z_{14}z_{23}|^{-1/2}\sum_{i,j=1}^3I_i^{\alpha}{\bar
I}_j^{{\bar \alpha}}F_{ij}(z,\bar z)
\end{equation}
where $\alpha$  denotes the ordered sequence
of  indices $\alpha_1,\alpha_2,\alpha_3,\alpha_4$, which label the
basis states of the four-dimensional representation of $osp(2|2)$, and where the invariant
tensors $I_1$, $I_2$ and $I_3$ are defined in equations
(\ref{osptensors}) together with similar equations for  ${\bar I}_1$,
${\bar I}_2$ and ${\bar I}_{3}$.
The
anharmonic ratio $z$ is defined as $z=z_{12}z_{34}/z_{14}z_{32}$ and
similarly for $\bar z$.
The functions $F_{ij}(z,{\bar z})$ are single-valued combinations of
the solutions to the $\widehat{osp}(2|2)_{-2}$ Knizhnik--Zamolodchikov
equations, which  are summarized by equations (\ref{fmatrix}) --- (\ref{lambda}). 
For example, focusing on the correlation function 
pertaining to the decomposition (\ref{Qsuper}):
\begin{gather}
\langle Q^{1\bar 1}(1) Q^{4\bar 4}(2) Q^{4\bar 4}(3) Q^{1\bar
1}(4)\rangle \sim |\Upsilon|^2\left[K(z)K(1-\bar z)+K(1-z)K(\bar z)\right].
\end{gather}
where here $\Upsilon=[z_{14}z_{23}\,z(1-z)]^{-1/4}$ and $K(z)$ and
$E(z)$ are the complete elliptic integrals. This is in agreement with
 both the replica result (\ref{qdiag}) and the strong disorder result
(\ref{strongrecast}) when the number of colours $N_C=2$.

\section{Dense Polymers and Twist Operators}
\label{dense}
It turns out that the supersymmetric description provides an extremely
economical and straightforward approach to the disordered Dirac
fermion problem. It follows from the work of Rasmussen \cite{Rassmu:free}, that the
action of the ${\widehat{osp}}(2|2)_{-2}$ WZNW model may be
represented as the direct sum of three simple theories:
\begin{equation}
\label{Rasmussendecomp}
\widehat{osp}(2|2)_{-2}=\widehat{su}(2)_{1}+\frac{1}{4\pi}\int d^2\xi\,
(\partial_\mu\varphi)^2+\int d^2\xi\,\epsilon_{ab}\,\partial_{\mu}\chi^a\partial^\mu\chi^b,
\end{equation}
where $\varphi$ is a non-compact bosonic field, $\chi^a$ is a
two-component symplectic fermion \cite{Kausch:Symplectic}, and $\epsilon_{ab}$ is a two-component
antisymmetric tensor. As expected in this supersymmetric theory, the bosonic sector ($c=2$) and the symplectic
fermions ($c=-2$) together yield a total central charge of zero.
We note that the presence of the $\widehat{su}(2)_{1}$ WZNW model in the decomposition
(\ref{Rasmussendecomp}) of the $\widehat{osp}(2|2)_{-2}$ WZNW model 
reflects an underlying  $\widehat{su}(2)_{-k/2}$ Ka{\v c}--Moody
subalgebra residing in the  $\widehat{osp}(2|2)_{k}$ algebra \cite{LeClair:Strong}.
The decomposition (\ref{Rasmussendecomp}) may be simplified even further by noting that the
$\widehat{su}(2)_1$ WZNW model admits the following free field
representation:
\begin{equation}
\widehat{su}(2)_1=\frac{1}{4\pi}\int d^2\xi\, (\partial_\mu\phi)^2,
\end{equation}
where $\phi$ is a compact free boson; note that our choice of
normalization is for latter convenience in equation (\ref{Qtwistcorr}). That is to say, the ${\widehat{osp}}(2|2)_{-2}$ WZNW model may be
represented as the sum of a compact bosonic field $\phi$ ($c=1$)
a non-compact bosonic field $\varphi$ ($c=1$) and a two-component
symplectic fermion ($c=-2$):
\begin{equation}
\label{ospdecomp}
\widehat{osp}(2|2)_{-2}=\int d^2\xi\, \left[\frac{1}{4\pi}(\partial_\mu\phi)^2 +\frac{1}{4\pi}(\partial_\mu\varphi)^2+\epsilon_{ab}\,\partial_{\mu}\chi^a\partial^\mu\chi^b\right].
\end{equation}
In view of the  decomposition (\ref{ospdecomp}) one anticipates a representation of the $\widehat{osp}(2|2)_{-2}$ primary
fields in terms of the primary fields of the models appearing on the
right-hand side of equation (\ref{ospdecomp}) --- namely vertex operators $e^{\alpha\phi}$ and $e^{\beta\varphi}$
and the primary fields of the $c=-2$ theory. As we shall discover in
sections \ref{osprecast} -- \ref{twistopcor} this is indeed possible. The non-unitary $c=-2$ minimal model  has
received a great deal of attention in recent years as a theory of
dense polymers \cite{Saleur:n=2,Ivashkevich:Polymers}, as a celebrated example of a
logarithmic conformal field theory
\cite{Gurarie:Log,Kausch:Curiosities,Gaberdiel:Rational,Gaberdiel:Local,Kausch:Symplectic,Kogan:Boundarylog},
and as a conformal ghost system \cite{FMS:confsuper}. The structure of
this theory is rather rich and is known to
consist of several sectors. As we shall see below, the so-called ${\Bbb
Z}_2$ twisted sector of the $c=-2$ theory will play a crucial r\^ole
here; this sector consists of a scalar primary field $\mu$ of conformal
dimension $-1/8$ and tensor primary fields $\nu^{\pm}$ of conformal
dimension $3/8$. The correlation functions and operator product
expansions  of these fields
have been studied in reference \cite{Saleur:n=2} and more fully in
reference \cite{Gaberdiel:Local}. In section \ref{osprecast} we shall
recast the four-point correlation functions  of the
$\widehat{osp}(2|2)_{-2}$ WZNW in a form
which facilitates the elucidation of the desired operator correspondence. In section \ref{twistops} we shall discuss the ${\Bbb Z}_2$ twist
operator correlation functions of the $c=-2$ model \cite{Gaberdiel:Local}. In section
\ref{twistopcor} we shall compare the
correlation functions (summarized in  Table \ref{regblocks} and Table
\ref{table:twist} of sections \ref{osprecast} and
\ref{twistops} respectively) and arrive at the aforementioned correspondence.

\subsection{The $\widehat{osp}(2|2)_{-2}$ WZNW model}
\label{osprecast}
In view of the large number of components of the generic $\widehat{osp}(2|2)_{-2}$ non-chiral four-point correlation
function and the rather cumbersome and opaque invariant tensors
(\ref{osptensors}) we shall refashion the correlator somewhat.\footnote{Restricting our attention to
four-point correlation functions pertaining to the four-dimensional $[0,1/2]$ representation
we have a total of $4^8$ components. Since the vast majority of these vanish
however, and many are related by symmetry, it is desirable to distill them further.} 
In particular the $c=-2$ operator correspondence will follow quite naturally. We study the four-point correlation function of the supersymmetric $Q$-matrix:
\begin{equation}
{\mathcal F}^{\alpha,\bar\alpha}(z_i,\bar
z_i)=\langle Q^{\alpha_1,\bar\alpha_1}(z_1,\bar
z_1) Q^{\alpha_2,\bar\alpha_2}(z_2,\bar
z_2) Q^{\alpha_1,\bar\alpha_3}(z_3,\bar
z_3) Q^{\alpha_4,\bar\alpha_4}(z_4,\bar z_4)\rangle,
\end{equation}
where on the left hand side we use the symbol $\alpha$ to denote the ordered
sequence of indices $\alpha_1,\alpha_2,\alpha_3,\alpha_4$. The
indices $\alpha_i$ assume the values $1,2,3,4$ and label the basis
states of the four-dimensional $[0,1/2]$ representation of $osp(2|2)$;
in the notation of \cite{Maassarani:non} states 1 and 4 are bosonic,
whilst states 2 and 3 are fermionic. As may be
seen from our more detailed studies in Appendix \ref{osp22k}, this
correlation function may be written in the (off-diagonal) form
\begin{equation}
\label{supercorr}
{\mathcal F}^{\alpha, \bar\alpha}(z_i,\bar z_{i}) = -{\mathcal
F}^{\alpha,\,(1)}(z_i){\mathcal F}^{\bar\alpha,\,(2)}(\bar z_i)-{\mathcal
F}^{\alpha,\,(2)}(z_i){\mathcal F}^{\bar\alpha,\,(1)}(\bar z_i),
\end{equation}
in which we have set the overall normalization of the four-point
function to minus unity for latter convenience, and where
\begin{equation}
\label{falpha}
{\mathcal
F}^{\alpha,\,(a)}(z_i)\equiv|z_{14}z_{23}|^{-1/4}\sum_{i=1}^3I_{i}^\alpha
F_{i}^{(a)}(z).
\end{equation}
Using the explicit form of the chiral blocks
$F_{i}^{(a)}(z)$ appearing in equations  (\ref{ellipticsolutions}),
(\ref{blocks1}) and  (\ref{blocks2}) together with the known transformation
properties of the invariant tensors --- see equations (\ref{jtensor})
and (\ref{jitensorcorr}) --- we deduce that
\begin{equation}
\label{ftrans}
{\mathcal F}^{\alpha,\,(2)}(z_{i})=-{\Tilde {\mathcal P}}{\mathcal
F}^{\Tilde\alpha,\,(1)}(\Tilde z_i).
\end{equation}
We use the tilde to denote the interchange of the coordinates or
indices $2$ and $3$, and ${\Tilde {\mathcal P}}$ to denote the
fermionic parity of this permutation; note that this permutation induces the
transformation $z\rightarrow 1-z$ in these functions. Substituting equation (\ref{ftrans}) into equation
(\ref{supercorr}) we obtain the result
\begin{equation}
{\mathcal F}^{\alpha, \bar\alpha}(z_i,\bar z_{i})=
{\Tilde{\Bar{\mathcal P}}}{\mathcal
F}^{\alpha,\,(1)}(z_i){\mathcal
F}^{\Tilde{\Bar\alpha},\,(1)}({\Tilde{\Bar z}}_i)+{\Tilde {\mathcal P}}{\mathcal
F}^{{\Tilde\alpha},\,(1)}({\Tilde z}_i){\mathcal F}^{\bar\alpha,\,(1)}(\bar z_i).\label{fullquota}
\end{equation}
That is to say, one may build the full quota of single-valued and
crossing symmetric  non-chiral correlation functions of the ${\widehat{osp}}(2|2)_{-2}$ WZNW
from our knowledge of the chiral functions ${\mathcal
F}^{\alpha,\,(1)}(z_{i})$ which display regular behaviour in
the vicinity of $z=0$ and logarithmic behaviour in the vicinity of
$z=1$. Using the explict results for the chiral blocks
(\ref{ellipticsolutions}), (\ref{blocks1}) and (\ref{blocks2}) together with
the invariant tensors (\ref{osptensors}) we
are able to gather these non-trivial functions in Table \ref{regblocks}. We
emphasize that the functions ${\mathcal
F}^{\alpha,\,(1)}(z_{i})$ are not simply the conformal blocks of the
${\widehat{osp}}(2|2)_{-2}$ WZNW model, but also embody the
non-trivial tensorial structure of $osp(2|2)$ --- see equation (\ref{falpha}).
\newlength{\LL} 
\begin{table}[tb]
\renewcommand{\arraystretch}{1.25}
\begin{center}
\begin{tabular}{|c||c|c|c|c||c|}
\hline
Sector & \multicolumn{4}{c||}{$\alpha$} & ${\mathcal
F}^{\alpha,\,(1)}(z_i)$\\
\hline\hline
\settowidth{\LL}{Bosonic}
\multirow{3}{\LL}{Bosonic} & \multicolumn{2}{c|}{1144} &
\multicolumn{2}{c||}{4411} & $(4\epsilon\gamma)^{-1}\,\Upsilon\,zK(z)$
\\
\cline{2-6}
 &  \multicolumn{2}{c|}{1414} & \multicolumn{2}{c||}{4141} &
$(4\epsilon\gamma)^{-1}\,\Upsilon\,(1-z)K(z)$ \\
\cline{2-6}
 &\multicolumn{2}{c|}{1441} & \multicolumn{2}{c||}{4114} &
$-(4\epsilon\gamma)^{-1}\,\Upsilon\,K(z)$\\
\hline\hline
\settowidth{\LL}{Fermionic}
\multirow{3}{\LL}{Fermionic} & \multicolumn{2}{c|}{2233} &
\multicolumn{2}{c||}{3322} & $(4\epsilon\gamma)\,\Upsilon\,\left[2E(z)-(2-z)K(z)\right]$
\\
\cline{2-6}
 &  \multicolumn{2}{c|}{2323} & \multicolumn{2}{c||}{3232} &
$(4\epsilon\gamma)\,\Upsilon\,\left[2E(z)-(1-z)K(z)\right]$ \\
\cline{2-6}
 &\multicolumn{2}{c|}{2332} & \multicolumn{2}{c||}{3223} &
$(4\epsilon\gamma)\,\Upsilon\,\left[2E(z)-K(z)\right]$\\
\hline\hline
\settowidth{\LL}{Mixed}
\multirow{6}{\LL}{Mixed} & 1234 & 1324 & 2143 & 3142 &
\settowidth{\LL}{$\pm \,\Upsilon\,[E(z)-(1-z)K(z)]$} \multirow{2}{\LL}{$\pm \,\Upsilon\,[E(z)-(1-z)K(z)]$}
\\
\cline{2-5}
& 2413 & 3412 & 4231 & 4321 & \\
\cline{2-6}
& 1243 & 1342 & 2134 & 3124 &
\settowidth{\LL}{$\pm\,\Upsilon\,[K(z)-E(z)]$} \multirow{2}{\LL}{$\pm\,\Upsilon\,[K(z)-E(z)]$}  \\
\cline{2-5}
& 2431 & 3421 & 4213 & 4312 & \\
\cline{2-6}
& 1423 & 1432 & 2314  & 3214 &
\settowidth{\LL}{$\pm\,\Upsilon\,E(z)$} \multirow{2}{\LL}{$\pm\,\Upsilon\,E(z)$}  \\
\cline{2-5}
& 2341 & 3241 & 4123 & 4132 & \\
\hline
\end{tabular}
\vspace{0.5cm}
\caption{Chiral correlation functions of the $\widehat{osp}(2|2)_{-2}$ WZNW model
displaying regular behaviour in the vicinity of $z=0$ and logarithmic behaviour in the vicinity of $z=1$; 
$\Upsilon=[z_{14}z_{23}\,z(1-z)]^{-1/4}$  and
$z=z_{12}z_{34}/z_{14}z_{32}$  --- see Appendix
\ref{osp22k} for further details.}
\label{regblocks}
\end{center}
\renewcommand{\arraystretch}{1}
\end{table}

\subsection{Twist Operator Correlation Functions}
\label{twistops}
The correlation functions of the twist operators $\mu$ and $\nu$ in the non-unitary
$c=-2$ minimal model have been extensively studied by Gaberdiel and
Kausch \cite{Gaberdiel:Local}. Their non-chiral four-point functions
are built from linear combinations of their chiral
counterparts so as to respect the stringent constraints of single-valuedness and
crossing symmetry. As may be seen from \S 4 of their work, the contributions
from either chiral sector are typically composed of two elliptic
integral solutions  posessing logarithmic branch cuts; one of these
solutions displays regular behaviour
in the vicinity of  $z=0$ and logarithmic behaviour in the vicinity of $z=1$, whilst the other displays
regular behaviour in the vicinity of $z=1$ and logarithmic behaviour
in the vicinity of $z=0$. As in section \ref{osprecast}, in order to establish
our correspondence (\ref{Qtwistcorr}) we find it
convenient to focus on the chiral contributions to the  four-point functions
which exhibit regular behaviour in the vicinity of $z=0$. The chiral
four-point functions most relevant to our discussion and displaying
such regular behaviour in
the vicinity of the origin are summarized in table \ref{table:twist}.

\begin{table}[tb]
\renewcommand{\arraystretch}{1.28}
\begin{center}
\begin{tabular}{|c||c||c|}
\hline
Sector & Correlator & Regular contribution \\
\hline\hline
$\mu$ & $\langle \mu\mu\mu\mu\rangle$ & $\Upsilon^{-1}\,K(z)$ \\
\hline\hline
\settowidth{\LL}{$\nu$} \multirow{3}{\LL}{$\nu$} &
$\langle\nu^{\pm}\nu^{\pm}\nu^{\mp}\nu^{\mp}\rangle$ &
$\Upsilon^3\,z\left[2E(z)-(2-z)K(z)\right]$ \\
\cline{2-3}
& $\langle\nu^{\pm}\nu^{\mp}\nu^{\pm}\nu^{\mp}\rangle$ & $\Upsilon^3\,(1-z)\,\left[2E(z)-(1-z)K(z)\right]$\\
\cline{2-3}
& $\langle\nu^{\pm}\nu^{\mp}\nu^{\mp}\nu^{\pm}\rangle$ & $\Upsilon^3\,\left[2E(z)-K(z)\right]$\\
\hline\hline
\settowidth{\LL}{Mixed} \multirow{6}{\LL}{Mixed} &
$\langle\mu\nu^\pm\nu^\mp\mu\rangle\times z_{14}^{-1/2}z_{23}^{+1/2}$ &
\settowidth{\LL}{$\pm\,\Upsilon\,[E(z)-(1-z)K(z)]$} \multirow{2}{\LL}{
$\pm\,\Upsilon\,[E(z)-(1-z)K(z)]$ 
}\\
& $\langle\nu^\pm\mu\mu\nu^\mp\rangle\times z_{14}^{+1/2}z_{23}^{-1/2}$ & \\
\cline{2-3}
& $\langle\mu\nu^\pm\mu\nu^\mp\rangle\times z_{13}^{-1/2}z_{24}^{+1/2}$ & 
\settowidth{\LL}{$\pm\Upsilon\,[K(z)-E(z)]$}
\multirow{2}{\LL}{$\pm\Upsilon\,[K(z)-E(z)]$}\\
& $\langle\nu^\pm\mu\nu^\mp\mu\rangle\times z_{13}^{+1/2}z_{24}^{-1/2}$ & \\
\cline{2-3}
& $\langle\mu\mu\nu^\pm\nu^\mp\rangle\times z_{12}^{-1/2}z_{34}^{+1/2}$ &
\settowidth{\LL}{$\pm\,\Upsilon \,E(z)$} \multirow{2}{\LL}{$\pm\,\Upsilon \,E(z)$}\\
& $\langle\nu^\pm\nu^\mp\mu\mu\rangle\times z_{12}^{+1/2}z_{34}^{-1/2}$ & \\
\hline
\end{tabular}
\vspace{0.5cm}
\caption{Chiral correlation functions of the ${\Bbb Z}_2$ twist
operators of $c=-2$  displaying regular behaviour
in the vicinity of $z=0$ and logarithmic behaviour in the vicinity of $z=1$; 
$\Upsilon=[z_{14}z_{23}\,z(1-z)]^{-1/4}$  and
$z=z_{12}z_{34}/z_{14}z_{32}$ --- see Gaberdiel and Kausch for further details
\cite{Gaberdiel:Local} and  note that our definition of
the anharmonic ratio differs from theirs.}
\label{table:twist}
\end{center}
\renewcommand{\arraystretch}{1}
\end{table}

\subsection{Twist Operator Correspondence}
\label{twistopcor}
Comparing our correlation functions for the
${\widehat{osp}}(2|2)_{-2}$ WZNW model (summarized in Table
\ref{regblocks}) with those of Gaberdiel and
Kausch for the ${\Bbb Z}_2$ twist fields of the  $c=-2$ minimal model \cite{Gaberdiel:Local}
(summarized in Table \ref{table:twist}) we see a clear correspondence
at the level of their elliptic integrals. In particular we see that
the bosonic and fermionic states of the $\widehat{osp}(2|2)_{-2}$ WZNW model are
naturally associated with the ${\Bbb Z}_2$ twist fields $\mu$
and $\nu_{\alpha}$ respectively. The remaining powers of
$z_{ij}$ which distinguish Table \ref{regblocks} from Table \ref{table:twist} are provided by bosonic vertex
operators as suggested by the decomposition (\ref{ospdecomp}). Explicitly,
we establish the following equivalence between the
chiral operators of the $\widehat{osp}(2|2)_{-2}$ WZNW model and the
`dressed' chiral operators of the $c=-2$ minimal model:\footnote{We
use the (overused and rather abused) symbol
$\sim$ to emphasize that we have derived the
correspondence from the chiral regular solutions and that
it is valid up to phase.}
\begin{equation}
\label{Qtwistcorr}
Q_1  \sim  {\mathcal A}^{-1}\,e^{i\phi}\,\mu, \quad
Q_2  \sim  {\mathcal A}\,e^{-\varphi}\,\nu^{-}, \quad
Q_3  \sim  {\mathcal A}\,e^{\varphi}\,\nu^{+}, \quad
Q_4  \sim  {\mathcal A}^{-1}\,e^{-i\phi}\,\mu,
\end{equation}
where ${\mathcal A}=(4\epsilon\gamma)^{1/4}$ is a free
parameter corresponding to the arbitrary relative normalizations of
the $su(2)$ doublet ($Q_{1}$, $Q_4$) and the two singlets ($Q_2$,
$Q_3$) in the four-dimensional representation of $osp(2|2)$
\cite{Scheurnert:ospm,Maassarani:non}. With the normalization chosen in equation (\ref{ospdecomp}) the compact bosonic exponents $e^{\pm
i\phi}$ have conformal dimension $h_{\phi}=1/4$, and the non-compact bosonic exponents $e^{\pm
\varphi}$ have $h_{\varphi}=-1/4$. It  is thus straightforward to
see that the dimensions add up to the correct value of $1/8$
--- equation (\ref{ospdim}) --- in the
decomposition (\ref{Qtwistcorr}). 

We have now highlighted the important
r\^ole  played by the $c=-2$ model in the chiral structure of the $\widehat{osp}(2|2)_{-2}$
WZNW model, and have provided a prescription for constructing the
non-chiral correlation functions (\ref{fullquota}). Although there
are many interesting and notable exceptions, we emphasize that in general, the non-chiral correlation
functions of the $\widehat{osp}(2|2)_{-2}$ WZNW are {\em not} simply
related to  {\em single} non-chiral correlation functions of the ${\Bbb Z}_2$ twist operators
dressed by non-chiral bosons. For example, it follows from our
solution of the $\widehat{osp}(2|2)_{-2}$ Knizhnik--Zamolodchikov
equations that
\begin{equation}
\langle Q^{2,\bar 2}(1)Q^{3,\bar 3}(2)Q^{2,\bar 2}(3)Q^{3,\bar
3}(4)\rangle={\mathcal C}\{[(1-z)K-2E][(1+\bar z)\Kbartilde-2\Ebartilde]+c.c.\},
\end{equation}
where ${\mathcal
C}=(16\epsilon^2\gamma^2)^2\,|z_{14}z_{23}|^{-1/2}\,\Lambda$, and $\Lambda$ is given by equation (\ref{lambda}). We note that this is {\em not} simply related to a {\em single}
correlation function of the non-chiral operators $\mu(z,\bar z)$ and
$\nu_{\alpha,\bar\alpha}(z,\bar z)$. That is to say, no {\em single} non-chiral
four-point function of the ${\Bbb Z}_2$ twist operators gives rise to
this particular single-valued combination of the elliptic
integrals. We invite the reader to verify this statement with the aid
of \S 4 of Gaberdiel and Kausch \cite{Gaberdiel:Local}. 
It may be interesting to study the non-chiral aspects of this
correspondence in greater detail.

\section{Convergence of Approaches}
\label{convergence}
In section \ref{localdos} we demonstrated that three different approaches to
the random Dirac fermion problem yield identical results for the
four-point correlation functions of the local density of states. In section \ref{dense} we
have further demonstrated that the supersymmetric approach inherits its
non-trivial logarithmic structure from the $c=-2$ non-unitary minimal model.
In this section we shall discuss the convergence of the these
approaches --- summarized in table \ref{Table:compare} --- and  the
emergence of the $c=-2$ theory in a little more detail. Indeed, we
shall point to a remarkable proliferation of $c=-2$ theories and effects. In table
\ref{Table:compare} we have highlighted the separation of the active (critical WZNW) degrees of
freedom from the passive (massive or decoupled Jacobian) degrees
of freedom in each approach.

\begin{table}[tb]
\renewcommand{\arraystretch}{1.25}
\begin{center}
\begin{tabular}{|c||c|c||c|c|}
\hline
 Approach & Fermions & Bosons & Active & Passive \\ \hline
Replicas & $N_C\times N_F$ & 0  & $\widehat{u}(1) + \widehat{su}(N_F)_{N_C}$ & $\widehat{su}(N_C)_{N_F}$\\ \hline
Strong Disorder& $N_C$ & 0 &  $\widehat{u}(1) +  \widehat{su}(N_C)_{1}
 + \widehat{su}(N_C)_{-2N_C}$ &  $(c=-2)^{N_C^2-1}$ \\ \hline
Supersymmetry & $2$ & $2$  & $\widehat{osp}(2|2)_{-2}$ & $\widehat{su}(2)_0$\\ \hline
\end{tabular}
\vspace{0.5cm}
\caption{Current algebra approaches used to investigate the critical behaviour of the
random non-Abelian Dirac fermion problem and the separation
of the active (critical WZNW) degrees of freedom
from the passive (massive or decoupled Jacobian) degrees of
freedom. The  central charges and the conformal
dimensions may be seen to add up correctly in each approach.}
\label{Table:compare}
\end{center}
\renewcommand{\arraystretch}{1}
\end{table}

An interesting aspect of this marriage of approaches is the
agreement between the weak coupling (replica
and supersymmetry) limit and the strong coupling limit. Although such
an agreement might have been anticipated purely on physical grounds,
its emergence is quite remarkable from a field-theoretic
perspective. At the very least, it is quite surprising that such
superficially different conformal field theories may nonetheless yield
equivalent results, albeit for a specific subset of physically
motivated correlation
functions. In particular, the resultant central charge of the active degrees
of freedom in the strong disorder approach differs from that in the
replica and supersymmetric approaches, namely zero. As we have discussed in section \ref{strongdisorder}
--- and indicate in table
\ref{Table:compare} ---  the active degrees of freedom in the strong
disorder limit may be expressed as the sum of three different models:
\begin{equation}
\label{strongdis}
S=\widehat{u}(1) + \widehat{su}(N_C)_{1} +
\widehat{su}(N_C)_{-2N_C}.
\end{equation}
where the first two models describe $N_C$ colours of free massless Dirac
fermions, and the remaining model encodes the non-trivial logarithmic
structure. We use the well known result for the central charge of
the $\widehat{g}_k$ WZNW model\cite{Knizam:current}
\begin{equation}
c=\frac{k\,\rm{dim}\,g}{k+g^\vee},
\end{equation}
where $g^\vee$ is the dual Coxeter number of the
algebra $g$ (equal to $N$ for $su(N)$)  together with the fact
that a free boson
$\widehat{u}(1)$ has central charge $1$, to find the resultant central charge of (\ref{strongdis}):
\begin{equation}
c=1+\frac{N_C^2-1}{1+N_C}+\frac{-2N_C(N_C^2-1)}{-2N_C+N_C}=2N_C^2+N_C-2\neq 0.
\end{equation}
That is to say, the active degrees of freedom in the strong disorder
approach  yield a positive central charge, in
contrast to the replica and supersymmetric degrees of freedom which
yield a net zero central charge.

As we have discused in section \ref{dense}, the
supersymmetric approach to the disordered Dirac fermion problem
reveals a hidden substructure which is inherited from the $c=-2$
minimal model. In view of the convergence of approaches outlined in
this paper it follows quite naturally that the active degrees of freedom in the
replica and strong disorder treatments also inherit non-trivial traits from the $c=-2$ model. In addition to the rather natural
proliferation of elliptic integrals in the conformal blocks presented
in this paper, we note
that  a relation between the $\widehat{su}(2)_{-4}$ WZNW model (which arises in the strong disorder treatment with $N_C=2$) and the
$c=-2$ theory has already been discussed in the string theory literature
\cite{Semikhatov:sl2}: the $\widehat{su}(2)_{-4}$ WZNW model is
cohomologically equivalent to an $N=4$ supersymmetric bosonic string
with $c=-2$ matter.\footnote{We note that this matter $+$ string theory
admits a description in terms of the $c=-2$ model dressed by  $c=28$
Liouville theory and $c=-26$ string ghosts. We remind the interested reader
that this (and other) Liouville LCFTs
\cite{Komudtsvelik,Bilal:Gravitationally,Bilal:Ongrav} emerged in a very natural way in
the closely related problem of prelocalization in disordered
conductors \cite{Komudtsvelik,Caux:exactmult}.}

A curious by-product of these
investigations is that the presence of the $c=-2$ theory may apparantly be
seen in the `passive' degrees of freedom outlined in table
\ref{Table:compare}. Indeed, a number of $c=-2$ models (arising as
Jacobians) explicitly decouple
in the strong disorder approach and play a passive (or spectator)
r\^ole in regards the LDOS.  A possible explanation for the closely related structure of the active and
passive theories indicated in table \ref{Table:compare} may arise from
the  requirement of the mutual cancellation of their logarithmic
singularities \cite{Bhaseen:legendre}. In particular we anticipate
a relationship between the
$\widehat{su}(2)_{0}$ WZNW model and the $c=-2$ theory and a
correspondence analogous to equation (\ref{Qtwistcorr}). We note that
the four-point functions of the $\widehat{su}(2)_{0}$ model have been
studied in references \cite{Caux:suzero,Kogan:origin}
and assume a simple form in terms of the complete elliptic
integrals \cite{Bhaseen:legendre}. The investigation  of the $\widehat{su}(2)_{0}$ WZNW model
will be continued in reference \cite{KoganNichols:suzero}.

As we close this section we comment briefly on the stability of the critical
point with respect to variations of the disorder distribution.
In the replica and supersymmetric approaches the gauge potential
disorder is coupled directly to the massive sector and it is natural
to assume that the critical theory is protected from such
variations. In the strong disorder approach, however, the
disorder variations are coupled directly to the critical $\widehat{su}(N_C)_{-2N_C}$ subsector, and
its stability is far from obvious. It is therefore interesting
to study the stability properties of this critical theory. A simple
perturbation of the critical theory that one might consider is the
deformation of the
WZNW action by its kinetic term. This operator has scaling dimension
zero and is strongly relevant. However, at level $k=-2N$ (or
more generally $-2g^\vee$) this operator commutes with all the
Ka{\v c}--Moody currents and therefore does not affect the correlation
functions --- see Appendix 6 of
reference \cite{Bhaseen:Towards} and note the different sign conventions for the level $k$.
In a more general framework, the WZNW models at level
$k=-2g^\vee$ arise quite
naturally in both two-dimensional ($c=0$) topological
field theories possessing a non-Abelian current algebra
\cite{Isidro:Topological}, and in the strong
disorder treatment of arbitrary WZNW models coupled to
random vector potentials \cite{Bernard:perturbed}. The critical level
arises from a BRST (Becchi--Rouet--Stora--Tyutin)
\cite{BRST:becchi,Tyutin} symmetry nilpotency condition and is
required for the coexistence of a Ka{\v c}--Moody algebra symmetry and a
topological algebra symmetry \cite{Isidro:Topological}.  We emphasize that
the level of the underlying Ka{\v c}--Moody algebra, the existence of a
topological algebra and the stability of
our strongly disordered theory are therefore intimately related.

\section{Conclusions}
\label{conclusions}
We briefly summarize here a number of our main results:
\begin{itemize}
\item{We have solved the random non-Abelian Dirac fermion problem by
means of three different non-perturbative procedures based on the
replica approach, supersymmetry, and in the limit of strong disorder. We have
demonstrated that this {\it m\'enage \`a
trois} of approaches yields identical results (for the four-point correlation
functions of the local density of states) in this relatively simple, but
quite non-trivial model of disorder.}

\item{We have emphasized the special r{\^o}le played by the level $k$ of the
Ka{\v c}--Moody algebra. As we have seen both here and in reference \cite{Bhaseen:legendre} the level
$k=-2$ is rather special for the $\widehat{osp}(2|2)_k$
WZNW model which arises in the supersymmetric treatment of the random Dirac fermion problem: an entire conformal block
decouples from the spectrum and the resulting theory is drastically
simplified.}

\item{We have established that the $c=-2$
minimal model plays an important r\^ole in the random non-Abelian 
Dirac fermion problem. We
have found a rather simple and
suggestive form for the supersymmetric critical action (\ref{ospdecomp}):
\begin{equation}
\notag
\widehat{osp}(2|2)_{-2}=u(1)+gl(1)+[c=-2].
\end{equation}
We have highlighted the relevance of the twist operators of the $c=-2$
theory in the disorder averaged correlation functions of
the Dirac fermion problem. The $c=-2$ symplectic fermions are
ubiquitous in the construction of supersymmetric sigma models
\cite{Efetov:super,Efetov:chaos}; this example indicates that this
sector may well be
responsible for many remarkable features of disordered critical points
including the presence of logarithmic operators \cite{Gurarie:Log}. We draw attention to the fact that the $c=-2$
theory emerges in the theoretical description of dense polymers
\cite{Saleur:n=2,Ivashkevich:Polymers}. We hope that this connection
may ultimately yield a more intuitive picture of the random
non-Abelian Dirac fermion problem
and other disordered critical points.}

\item{We have argued for the stability of the critical point with
respect to variations in the disorder, and noted that the level of the underlying Ka{\v c}--Moody algebra, the existence of a
topological algebra and the stability of
our strongly disordered theory are rather intimately linked. It is
an interesting open question whether such
considerations may yield Ka{\v c}--Moody level selection
mechanisms in models displaying lines of critical points \cite{Zirnbauer:integer}.}
\end{itemize}

\section{Acknowledgements}
We gratefully acknowledge valuable conversations with V. Gurarie and
C. P{\'e}pin. Part of this work was carried out at the Max Planck
Institute in Dresden and three of us (MJB, JSC and AMT)
acknowledge its hospitality. MJB would also like to thank EPSRC for
financial support. 
\vspace*{0.5cm}

{\it Note added in proof}: whilst this paper was being completed we became aware of a preprint by A. W. W. Ludwig \cite{Ludwig:c=-2}
in which one of our results, namely the equivalence (\ref{ospdecomp}), was
established.
\newpage
\appendix
\section{The \sunk WZNW model}
\label{sunapp}
In this appendix we consider the appearance of logarithms in the
correlation functions of the $\widehat{\mathit{su}}(N)_k$ WZNW model. Following $\S 4$ of Knizhnik and Zamolodchikov
\cite{Knizam:current} we compute the four-point functions 
\begin{equation}
{\mathcal F}^{{\boldsymbol \alpha},\bar{\boldsymbol\alpha}}(z_i,\bar
z_i)=\langle g(z_1,\bar
z_1)g^\dagger (z_2,\bar
z_2)g^\dagger (z_3,\bar
z_3)g(z_4,\bar z_4)\rangle,
\end{equation}
of the field $g(z_i,{\bar z_i})=g^{\alpha_i,\bar\alpha_i}(z_i,\bar z_i)$
transforming in the fundamental representation of SU(N)$\times$SU(N). We use the symbol ${\boldsymbol \alpha}$ to denote the ordered
sequence of indices $\alpha_1,\alpha_2,\alpha_3,\alpha_4$. Global conformal invariance restricts this correlation function to
have the form
\begin{equation}
\label{globalfourptsu}
{\mathcal F}^{{\boldsymbol\alpha},\bar{\boldsymbol\alpha}}(z_i,\bar
z_i)=(z_{14}z_{23}\bar z_{14}\bar z_{23})^{-2h}F^{{\boldsymbol\alpha},\bar{\boldsymbol\alpha}}(z,\bar z)
\end{equation}
where $z$ and $\bar z$ are the anharmonic ratios
\begin{equation}
\label{anharm}
z=\frac{z_{12}z_{34}}{z_{14}z_{32}}, \quad \bar z=\frac{\bar
z_{12}\bar z_{34}}{\bar z_{14}\bar z_{32}}
\end{equation}
and the conformal dimension $h$ of the field $g$ is
\begin{equation}
\label{dimensionsu}
h=\frac{N^2-1}{2N(N+k)}
\end{equation}
The correlation function (\ref{globalfourptsu}) admits the su$(N)$$\times$su$(N)$
invariant decomposition
\begin{eqnarray}\label{fijsu}
F^{{\boldsymbol\alpha},\bar{\boldsymbol\alpha}}(z,\bar
z)=\sum_{ij=1}^{2}I_{i}^{\boldsymbol\alpha}{\bar I}_j^{\boldsymbol{\bar\alpha}} F_{ij}(z,\bar z)
\end{eqnarray}
where the invariant tensors $I_1$ and $I_2$ are defined as 
\begin{equation}
I_1^{{\boldsymbol\alpha}}=\delta^{\alpha_1,\alpha_2}\delta^{\alpha_3,\alpha_4}, \quad I_2^{{\boldsymbol\alpha}}=\delta^{\alpha_1,\alpha_3}\delta^{\alpha_2,\alpha_4}
\end{equation}
with similar equations for  ${\bar I}_1$ and ${\bar I}_2$. The four scalar functions
$F_{ij}$ satisfy the coupled first-order differential equations \cite{Knizam:current}
\begin{equation}
\frac{dF}{dz}=\left[\frac{1}{z}P+\frac{1}{z-1}Q\right]F, \quad
\mbox{where} \quad  F=\begin{pmatrix} F_{1j} \\ F_{2j}
\end{pmatrix} \quad \forall j 
\end{equation}
where the matrices $P$ and $Q$ are given by
\begin{equation}
\label{PQSUN}
P=-\frac{1}{N(N+k)}\begin{pmatrix} N^2-1 & N \\ 0 & -1 \end{pmatrix},
\quad Q=-\frac{1}{N(N+k)}\begin{pmatrix} -1 & 0 \\ N & N^2-1 \end{pmatrix}
\end{equation}
There are similar equations for the antiholomorphic
dependence. Suppressing the antiholomorphic index $j$ from the
functions $F_{1j}$ and $F_{2j}$, one may obtain the second-order
differential equation satisfied by $F_1(z)$:
\begin{align}
\begin{split}
& N^2(N+k)^2z^2(1-z)^2 F_1^{\prime\prime}(z)- \\
& \hspace{1.5cm} N(N+k)z(1-z)\left[2-N(2N+k)+(3N^2+Nk-4)z\right]F_1^\prime(z) + \\
& \hspace{3cm}
\left[1-N^2-(N^4-6N^2-Nk+4)z+(N^4-5N^2+4)z^2\right]F_1(z)=0.
\end{split} 
\end{align}
The corresponding $F_2(z)$  may be obtained from the $F_1(z)$
solutions:
\begin{align}
\begin{split}
\label{f2sun}
F_2(z)=-(N+k)zF_1^\prime(z)+N^{-1}(1-z)^{-1}\left[1-N^2+(N^2-2)z\right]F_1(z).
\end{split}
\end{align}
Upon the change of variables 
\begin{equation}
\label{chiralsun}
F_1(z)=z^{-N/(N+k)}[z(1-z)]^{1/N(N+k)}G_1(z)
\end{equation}
one obtains an equation of the hypergeometric form:
\begin{equation}
\label{hypersu}
z(1-z)G_1^{\prime\prime}+\left[\gamma-(\alpha+\beta+1)z\right]G_1^\prime-\alpha\beta
G_1=0.
\end{equation}
where $\alpha=-1/(N+k)$, $\beta=1/(N+k)$ and $\gamma=k/(N+k)$. For
non-integer values of $\gamma$ the solutions of these equations are
given by Knizhnik and Zamolodchikov --- see equations (4.10a) and
(4.10b) of \cite{Knizam:current}. For integer values of
$\gamma$, however, the solution involves $\it logarithms$ --- the roots of the indicial equation corresponding to
(\ref{hypersu}) (namely $0$ and $1-\gamma$) differ by an
integer. 
Of particular importance in the study of the disordered Dirac
fermion problem are the cases $N\rightarrow 0$
($\gamma=1$) appearing in the replica treatment (section \ref{replicaapproach}) and $k=-2N$ ($\gamma=2$) appearing in the strong
disorder treatment (section \ref{strongdisorder}). We consider both of
these cases below.

\subsection{The $\widehat{u}(N)_k$ WZNW Model as $N\rightarrow 0$}
\label{unkwznw}
As may be seen from equation (\ref{dimensionsu}) the conformal dimensions of the
$\widehat{\mathit{su}}(N)_k$ WZNW model {\it diverge} as $N\rightarrow 0$, as do some of
the prefactors in the chiral blocks (\ref{chiralsun}); the $N\rightarrow 0$ limit of the $\widehat{\mathit{su}}(N)_k$ WZNW model {\it
alone} fails to  yield  a well defined CFT. However, the replica
approach to the random Dirac fermion problem instructs us to consider
the $\widehat{u}(N)_k$ WZNW model. As we shall demonstrate below, the
 $\widehat{u}(N)_k$ WZNW model {\it does} have a well defined $N\rightarrow 0$
limit. 

An arbitrary element $u$, of the group U(N), may be expressed as an
element of
SU(N), $g$, multiplied by a phase: $u=e^{i\alpha\varphi}g$. Using the
Polyakov--Wiegmann identity for the WZNW model \cite{Poly:multi}
\begin{equation}
\label{polywieg}
W_k[ab]=W_k[a]+W_k[b]+\frac{k}{2\pi}\int
d^2\xi\,\tr^\prime(a^{-1}\bar\partial ab\partial b^{-1})
\end{equation}
with $a=e^{i\alpha\varphi}$ and $b=g$ one obtains\footnote{We have used the facts that the current $g\partial g^{-1}$
residing in the su(N) {\it algebra} is traceless, thereby eliminating
the second term in (\ref{polywieg}), that the WZNW term (\ref{WZNW}) vanishes
for the U(1) element $e^{i\alpha\varphi}$, and that $\tr^\prime\equiv
2\tr$.} 
\begin{eqnarray}
W_k[u] = W_k[g]+\frac{\alpha^2Nk}{8\pi}\int d^2\xi\, \partial_\mu\varphi\partial^\mu\varphi
\end{eqnarray}
The $\widehat{u}(N)_k$ WZNW model is therefore seen to be the $\widehat{\mathit{su}}(N)_k$ WZNW model augmented by a free scalar field of {\it prescribed}
normalization. The conformal dimension of the field $e^{i\alpha\varphi}$ is {\it fixed} by this normalization
to be\footnote{For a free scalar field governed by the action
$S=\frac{1}{2}g\int d^2\xi \, \partial_\mu\varphi\partial^\mu\varphi$,
the field $e^{i\alpha\varphi}$ has $h_\alpha=\alpha^2/8\pi
g$ --- see for example equations (5.73) and (6.60) of
\cite{Francesco:CFT}.} $h_\alpha=1/(2Nk)$. The conformal dimension of
the composite field $u=e^{i\alpha\varphi}g$ is thus
\begin{equation}
\label{repdim}
h_{\widehat{u}(N)_k}=\frac{1}{2Nk}+\frac{N^2-1}{2N(N+k)}\xrightarrow[N\rightarrow
0]{}
\frac{1}{2k^2}
\end{equation}
and is seen to have the finite replica limit $1/2k^2$. In addition, the
(non-vanishing)
holomorphic  four-point correlation function of the U(1) fields
${\mathcal V}_\alpha=e^{i\alpha\varphi}$ is given by
\begin{equation}
\label{u1corr}
\langle{\mathcal V}_\alpha(1){\mathcal V}_{-\alpha}(2){\mathcal
V}_{-\alpha}(3){\mathcal V}_{\alpha}(4)\rangle  = \left(z_{14}z_{32})^{-1/Nk}[z(1-z)\right]^{-1/Nk}
\end{equation}
The divergences occuring in the SU(N)$_k$ results (\ref{globalfourptsu}) and
(\ref{chiralsun}) as $N\rightarrow 0$ are seen to be compensated by those of
the U(1) phase (\ref{u1corr}). In the replica ($N\rightarrow 0$) limit
the hypergeometric equation (\ref{hypersu}) reads
\begin{equation}
z(1-z)G_{1}^{\prime\prime}+(1-z)G_1^\prime+k^{-2}G_1=0
\end{equation}
and one is able to find the $N\rightarrow 0$ limit of the $\widehat{u}(N)_k$ conformal blocks:
\begin{subequations}
\label{replicablocks}
\begin{eqnarray}
F_{1}^{(1)} & = &\gamma\,\ _2F_1[-\tfrac{1}{k},\tfrac{1}{k};1;z] \\
F_{1}^{(2)} & = & \gamma\,(1-z) \ _2F_1[1-\tfrac{1}{k},1+\tfrac{1}{k};2;1-z] \\
F_{2}^{(1)} & = & \gamma\,\frac{z}{k}\ _2F_1[1-\tfrac{1}{k},1+\tfrac{1}{k};2;z] \\
F_{2}^{(2)} & = & \gamma\, k\ _2F_1[-\tfrac{1}{k},\tfrac{1}{k};1;1-z]
\end{eqnarray}
\end{subequations}
where $\gamma=[z(1-z)]^{-1/k^2}$. We find it convenient to introduce generalizations of the complete
elliptic integrals of the first and second kind \cite{bhaseen:zn}:
\begin{subequations}
\label{genelliptic}
\begin{eqnarray}
K_k(z) & \equiv & \frac{\pi}{k}\ _2F_1[1-\tfrac{1}{k},\tfrac{1}{k};1;z] \\
E_k(z) & \equiv & \frac{\pi}{k}\ _2F_1[-\tfrac{1}{k},\tfrac{1}{k};1;z] 
\end{eqnarray}
\end{subequations}
In terms of these functions (and rescaling) the $N\rightarrow 0$ $\widehat{u}(N)_k$
conformal blocks read:
\begin{subequations}
\label{replicablocks2}
\begin{eqnarray}
F_{1}^{(1)} & = & \gamma\,E_k(z)\\
F_{1}^{(2)} & = & \gamma\,k\,\left[K_k(1-z)-E_k(1-z)\right] \\
F_{2}^{(1)} & = &  \gamma\,\left[K_k(z)-E_k(z)\right]\\
F_{2}^{(2)} & = & \gamma\,k\,E_k(1-z)
\end{eqnarray}
\end{subequations}
The full $U(N)\times U(N)$ invariant correlation function is built
from single-valued combinations of these conformal blocks; it is straightforward to see that one can only have
\begin{equation}
\label{repsingle}
F_{ij}(z, \bar z) = X_{12}\left[F_i^{(1)}(z)F_j^{(2)}({\bar z})+F_i^{(2)}(z)F_j^{(1)}({\bar z})\right]
\end{equation}
Substituting the explicit forms (\ref{replicablocks2}) into (\ref{repsingle}) one obtains
\begin{subequations}
\label{replicafij}
\begin{align}
F_{11}&=X_{12}\,k\,|\gamma|^2\,\left[E_{k}(\Tilde{\Bar K}_{k}-\Tilde{\Bar E}_{k})+\Bar{E}_{k}(\Tilde{K}_{k}-\Tilde{E}_{k})\right] \\
F_{12}& =
X_{12}\,k\,|\gamma|^2\,\left[E_{k}\Tilde{\Bar{E}}_{k}+(\Tilde{K}_{k}-\Tilde{E}_{k})(\Bar{K}_{k}-\Bar{E}_{k})\right]\\
F_{21}& =
X_{12}\,k\,|\gamma|^2\,\left[{\Bar E}_{k}\Tilde{E}_{k}+(\Tilde{\Bar
K}_{k}-\Tilde{\Bar E}_{k})(K_{k}-E_{k})\right]\\
F_{22}&=X_{12}\,k\,|\gamma|^2\,\left[{\Tilde E}_{k}({\Bar K}_{k}-{\Bar E}_{k})+\Tilde{\Bar{E}}_{k}(K_{k}-E_{k})\right]
\end{align}
\end{subequations}
where we have adopted the notation that  ${\Tilde f}(z)\equiv f(1-z)$, ${\bar f}(z)\equiv f({\Bar
z})$ and ${\Tilde {\Bar f}}(z)\equiv f(1-{\Bar z})$ for an arbitrary
function $f$.

\subsection{The $\widehat{\mathit{su}}(N)_{-2N}$ WZNW Model}
\label{sunminus2n}
As may be seen from equations (\ref{chiralsun}) and (\ref{hypersu}) with $k=-2N$, the $F_1(z)$
chiral blocks for the $\widehat{\mathit{su}}(N)_{-2N}$ WZNW Model are given by
\begin{equation}
F_1(z)=z\left[z(1-z)\right]^{-1/N^2}\,G_1(z)
\end{equation}
where $G_1$ satisfies the hypergeometric equation with $\alpha=1/N$,
$\beta=-1/N$, $\gamma=2$:
\begin{equation}
\label{stronghyper}
z(1-z)G_1^{\prime\prime}+(2-z)G_{1}^{\prime}+N^{-2}G_1=0
\end{equation}
Solving this equation and applying the relation (\ref{f2sun}) enables
one to obtain the full set of $\widehat{\mathit{su}}(N)_{-2N}$ chiral blocks:\footnote{We note that
(\ref{f12str}) may be obtained by substituting
$G_1(z)=z^{-1}(1-z)^2H_1(z)$ into (\ref{stronghyper}) and replacing $z$
by $1-z$ --- this yields a hypergeometric equation. The result
(\ref{f21str}) may be obtained by straightforward application of (\ref{f2sun}) to (\ref{f11str}). In deriving
(\ref{f22str}) one may utilize the result $(N^2-1)z(1-z)\
_2F_1[2-\tfrac{1}{N},2+\tfrac{1}{N};4;z]+3N^2(2-z)\
_2F_1[1-\tfrac{1}{N},1+\tfrac{1}{N};3;z]=6N^2\
_2F_1[-\tfrac{1}{N},\tfrac{1}{N};2;z]$ which is easily verified by means of the Gauss
recursion relations.}
\begin{subequations}
\label{strongblocks}
\begin{eqnarray}
F_{1}^{(1)} & = & \Lambda \,z\ _2F_1[-\tfrac{1}{N},\tfrac{1}{N};2;z] \label{f11str}\\
F_{1}^{(2)} & = & \Lambda \,(1-z)^2 \ _2F_1[1-\tfrac{1}{N},1+\tfrac{1}{N};3;1-z] \label{f12str}\\
F_{2}^{(1)} & = & -\frac{\Lambda z^2}{2N}\ _2F_1[1-\tfrac{1}{N},1+\tfrac{1}{N};3;z] \label{f21str}\\
F_{2}^{(2)} & = & -2N\Lambda\,(1-z)\ _2F_1[-\tfrac{1}{N},\tfrac{1}{N};2;1-z]\label{f22str}
\end{eqnarray}
\end{subequations}
where $\Lambda=\left[z(1-z)\right]^{-1/N^2}$.  In terms of generalized
elliptic integrals the $\widehat{\mathit{su}}(N)_{-2N}$ chiral blocks read:
\begin{subequations}
\label{strongblocksellip}
\begin{eqnarray}
F_{1}^{(1)} & = & c\Lambda
\left[\left[1+(N-1)z\right]E_N-(1-z)K_N\right] \label{f11ell}\\
F_{1}^{(2)} & = & 2Nc\Lambda \left[\left[1+(N-1)z\right]{\Tilde
E}_N-Nz{\Tilde K}_N\right] \label{f12ell} \\
F_{2}^{(1)} & = & -c\Lambda \left[\left[N-(N-1)z\right]E_N-N(1-z)K_N\right]  \label{f21ell}\\
F_{2}^{(2)} & = & -2Nc\Lambda
\left[\left[N-(N-1)z\right]{\Tilde E}_N-z{\Tilde K}_N\right]\label{f22ell}
\end{eqnarray}
\end{subequations}
where $c=N^2/[\pi(N^2-1)]$. We have used the Gauss recursion formula GR 9.137(13) \cite{Gradshteyn}
\begin{equation}
\begin{split}
&\gamma[\alpha-(\gamma-\beta)z]\ _2F_1[\alpha,\beta;\gamma;z]-\alpha\gamma(1-z)
\ _2F_1[\alpha+1,\beta;\gamma;z] \\ & \hspace{1.5cm}+(\gamma-\alpha)(\gamma-\beta)z\ _2F_1[\alpha,\beta;\gamma+1;z]=0
\end{split}
\end{equation}
with $\alpha=-1/N$, $\beta=1/N$, $\gamma=1$ to obtain (\ref{f11ell}),
and GR 9.137(6) \cite{Gradshteyn} 
\begin{equation}
\begin{split}
&\gamma(\gamma+1)\ _2F_1[\alpha,\beta;\gamma;z]-\gamma(\gamma+1)\
_2F_1[\alpha,\beta;\gamma+1;z] \\ & \hspace{1cm}-\alpha\beta z\ _2F_1[\alpha+1,\beta+1;\gamma+2;z]=0
\end{split}
\end{equation}
with  $\alpha=-1/N$, $\beta=1/N$, $\gamma=1$ to obtain (\ref{f12ell}).
One may obtain (\ref{f21ell}) and (\ref{f22ell}) by noting that
$F_2^{(1)}(z)=-F_1^{(2)}(1-z)/2N$ and
$F_2^{(2)}(z)=-2NF_1^{(1)}(1-z)$ as follows from
(\ref{strongblocks}). In particular we note that
\begin{subequations}
\label{strongtraceblocks}
\begin{eqnarray}
N^2F_1^{(1)}+NF_2^{(1)} & = & c^\prime \Lambda z E_N \\
NF_1^{(1)}+N^2F_2^{(1)} & = & c^\prime \Lambda (1-z) [K_N-E_N]\\
N^2F_1^{(2)}+NF_2^{(2)} & = & -2Nc^\prime \Lambda z [{\Tilde
K}_N-{\Tilde E}_N]\\
NF_1^{(2)}+N^2F_2^{(2)} & = & -2Nc^\prime \Lambda (1-z) {\Tilde E}_N
\end{eqnarray}
\end{subequations}
where $c^\prime\equiv cN(N^2-1)$ is simply another constant.

\section{The \ospk WZNW model}
\label{osp22k}
\subsection{Representations of $osp(2|2)$}
The Lie superalgebra osp$(2|2)$ $\sim$ spl$(2|1)$ consists of four bosonic
generators $Q_{3}$, $Q_{+}$, $Q_{-}$, $B$, and four fermionic
generators $W_{+}$, $W_{-}$, $V_{+}$, $V_{-}$. The bosonic subalgebra is $sl(2)\oplus u(1)$:
\begin{equation}
[Q{_3},Q_{\pm}]=\pm Q_{\pm}, \quad [Q_{+}, Q_{-}]=2Q_{3}, \quad
[B,Q_{\pm}]=0, \quad [B,Q_{3}]=0 
\end{equation}
and the eigenvalues of $Q_{3}$ and $B$ (called isospin and baryon number
respectively) are used to classify the basis states of finite-dimensional
representations \cite{Scheurnert:ospm,Scheurnert:ospm2}.  A
representation $[b,q]$ contains at most four multiplets of states:
\begin{equation}
|b,q\rangle, \quad |b+\tfrac{1}{2},q-\tfrac{1}{2}\rangle, \quad
|b-\tfrac{1}{2},q-\tfrac{1}{2}\rangle, \quad |b,q-1\rangle,
\end{equation}
the states within a given multiplet $|b,q\rangle$ being labelled
by their third component of isospin $|b,q,q_3\rangle$. In particular
the $[0,\tfrac{1}{2}]$ representation is four-dimensional and contains
a doublet and two singlets (the multiplet $|b,q-1\rangle$ being absent):
\begin{equation}
|0,\tfrac{1}{2},\tfrac{1}{2}\rangle, \quad 
|0,\tfrac{1}{2},-\tfrac{1}{2}\rangle, \quad |\tfrac{1}{2},0,0\rangle,
\quad |-\tfrac{1}{2},0,0\rangle.
\end{equation}
 Following \cite{Maassarani:non} we
denote these states as $|1\rangle$, $|4\rangle$, $|3\rangle$ and
$|2\rangle$ respectively. In addition to their isospin and baryon
number, the basis states carry a grading denoted by
$\varepsilon=\pm 1$. States  $|1\rangle$ and  $|4\rangle$ are
even/bosonic and have $\varepsilon=0$, whilst states $|2\rangle$ and
$|3\rangle$ are odd/fermionic and have $\varepsilon=1$. On this basis
of states one may construct $4\times4$ matrix representation
$[0,\tfrac{1}{2}]$ of the generators of osp$(2|2)$.

\subsection{Knizhhnik--Zamolodchikov Equations}
Let us study the four-point function of the supersymmetric $Q$-matrix:
\begin{equation}
\label{superfour}
{\mathcal F}^{{\boldsymbol \alpha},\bar{\boldsymbol\alpha}}(z_i,\bar
z_i)=\langle Q^{\alpha_1,\bar\alpha_1}(z_1,\bar
z_1) Q^{\alpha_2,\bar\alpha_2}(z_2,\bar
z_2) Q^{\alpha_1,\bar\alpha_3}(z_3,\bar
z_3) Q^{\alpha_4,\bar\alpha_4}(z_4,\bar z_4)\rangle,
\end{equation}
where on the left hand side we use the symbol ${\boldsymbol \alpha}$ to denote the ordered
sequence of indices $\alpha_1,\alpha_2,\alpha_3,\alpha_4$. Global conformal invariance restricts this correlation function to
have the form
\begin{equation}
\label{globalfourpt}
{\mathcal F}^{{\boldsymbol\alpha},\bar{\boldsymbol\alpha}}(z_i,\bar
z_i)=(z_{14}z_{23}\bar z_{14}\bar z_{23})^{-2h}F^{{\boldsymbol\alpha},\bar{\boldsymbol\alpha}}(z,\bar z)
\end{equation}
where $z$ and $\bar z$ are the anharmonic ratios
\begin{equation}
z=\frac{z_{12}z_{34}}{z_{14}z_{32}}, \quad \bar z=\frac{\bar
z_{12}\bar z_{34}}{\bar z_{14}\bar z_{32}}
\end{equation}
and the conformal dimension $h$ of the field $\phi$ is
\begin{equation}
\label{ospdim}
h = \frac{1}{4 - 2k}
\end{equation}
The correlation function (\ref{globalfourpt}) has the osp$(2|2)$$\times$osp$(2|2)$
invariant decomposition
\begin{eqnarray}\label{fij}
F^{{\boldsymbol\alpha},\bar{\boldsymbol\alpha}}(z,\bar
z)=\sum_{ij=1}^{3}I_{i}^{\boldsymbol\alpha}{\bar I}_j^{\boldsymbol{\bar\alpha}} F_{ij}(z,\bar z)
\end{eqnarray}
The tensors $I$ and $\bar I$ are given in appendix A of
 \cite{Maassarani:non}:
\begin{subequations}
\label{osptensors}
\begin{eqnarray}
\label{ione}
I_1 & = & (1144)+(1234)4  \ep\ga+(1324)4  \ep\ga-(1414)+(2143)4  \ep\ga 
\nonumber\\&+&(2233)16 \ep^2\ga^2+
(2323)16 \ep^2\ga^2-(2413)4  \ep\ga+(3142)4  \ep\ga 
+(3232)16 \ep^2\ga^2 \nonumber \\
&+& (3322)16 \ep^2\ga^2 -(3412)4  \ep\ga-(4141)
-(4231)4  \ep\ga-(4321)4  \ep\ga+(4411)\\
& & \nonumber \\
\label{itwo}
I_2&=& (1234)4  \ep\ga-(1243)4  \ep\ga+(1324)4  \ep\ga-(1342)4  
\ep\ga \nonumber\\
&-& (1414)+(1441)-(2134)4  \ep\ga +(2143)4  \ep\ga+(2233)32 \ep^2\ga^2
+(2323)16 \ep^2\ga^2 \nonumber\\
&+& (2332)16 \ep^2\ga^2-(2413)4  \ep\ga
+(2431)4  \ep\ga-(3124)4  \ep\ga+(3142)4  \ep\ga \nonumber\\
&+& (3223)16 \ep^2\ga^2+(3232)16 \ep^2\ga^2
+(3322)32 \ep^2\ga^2-(3412)4  \ep\ga + (3421)4  \ep\ga \nonumber\\
&+&(4114) - (4141)+(4213)4  \ep\ga
-(4231)4  \ep\ga+(4312)4  \ep\ga-(4321)4  \ep\ga\\
&  & \nonumber \\
\label{ithree}
I_3&=&(1234)-(1243)+(1324)-(1342)+(1423)+(1432)-(2134)+(2143)
\nonumber\\ 
&+&(2233)8  \ep\ga +(2314)+(2323)8  \ep\ga+(2332)8  \ep\ga
-(2341)-(2413)+(2431) \nonumber \\
&-& (3124)+(3142)+(3214)+(3223)8  \ep\ga+(3232)8  \ep\ga-(3241)
+(3322)8  \ep\ga \nonumber \\
&-& (3412)+(3421)-(4123)-(4132)+(4213)-(4231)+(4312)-(4321)
\end{eqnarray}
\end{subequations}
and the nine scalar functions
$F_{ij}$ satisfy the coupled first-order differential equations \cite{Maassarani:non}
\begin{equation}
x\frac{dF}{dz}=\left[\frac{1}{z}P+\frac{1}{z-1}Q\right]F, \quad
\mbox{where} \quad  F=\begin{pmatrix} F_{1j} \\ F_{2j} \\F_{3j}
\end{pmatrix} \quad \forall j 
\end{equation}
There are similar equations for the antiholomorphic
dependence. Suppressing the antiholomorphic index $j$ from the
functions $F_{1j}\cdots F_{3j}$, one may reduce this first-order matrix
differential equation to the following set of equations \cite{Maassarani:non}
\begin{subequations}
\begin{align}
\begin{split}
& x^3z^3(1-z)^3\frac{d^3F_3(z)}{dz^3}+x^2(1+2x)z^2(1-z)^2(1-2z)\frac{d^2F_3(z)}{dz^2} + \\ 
& \hspace{3cm} xz(1-z)[-1-x+2xz-2x(2+x)z(1-z)]\frac{dF_3(z)}{dz} + \\
& \hspace{7cm} [-1-x+2z+2xz(1-z)]F_3(z)=0
\end{split} \label{f3} \\
\begin{split}
F_2(z) = -\frac{1}{4\epsilon\gamma
xz(1-z)}\left[x^2D^2F_3(z)+2x(1-z)DF_3(z)+(1-2z)F_3(z)\right] 
\end{split}\label{f2} \\
\begin{split}
F_1(z)=\frac{1}{4\epsilon\gamma}\left[xDF_3(z)-F_3(z)\right]+(z-2)F_2(z)
\end{split}\label{f1}
\end{align}
\end{subequations}
where $D=z(1-z)d/dz$. Equation (\ref{f3}) has three independent solutions which we denote
$F_{3}^{(a)}$ where  $a=1,2,3$.  Equations (\ref{f2}) and
(\ref{f1}) yield the corresponding solutions $F_2^{(a)}$ and
$F_1^{(a)}$. The nine scalar functions $F_{ij}$ appearing in equation
(\ref{fij}) may be expressed as a linear combination of these nine
functions (the so-called chiral/current blocks.)
\begin{equation}
F_{ij}(z, \bar z)=\sum_{a,b=1}^3X_{ab}F_i^{(a)}(z)F_j^{(b)}(\bar z)
\end{equation}
The values of the coefficients $X_{ab}$ are determined by single
valuedness (monodromy invariance) and crossing symmetry  to be
discussed later.
\subsubsection{Factorization at Level $k=1$}
At level $k=1$ ($x=1$) the Knizhnik--Zamolodchikov
equation (\ref{f3}) takes the form
 \begin{equation}
z^3(1-z)F_3^{'''}+3z^2(1-2z)F_3^{''}-2z(1+3z)F_3^{'}-2F_3=0
\end{equation}
This may be factorized to read
\begin{equation}
\label{factorizek1}
{\mathcal D}^{(2)}{\mathcal D}^{(1)}F_3=0
\end{equation}
where
\begin{eqnarray}
{\mathcal D}^{(1)} & = & z\frac{d}{dz}+1 \\
{\mathcal D}^{(2)} & = & z^2(1-z)\frac{d^2}{dz^2}-3z^2\frac{d}{dz}-2
\end{eqnarray}
It is straightforward to see that the equation ${\mathcal
D}^{(1)}F_3=0$, and thus (\ref{factorizek1}) has the solution
\begin{equation}
F_{3}^{(1)}=\frac{1}{z} 
\end{equation}
Applying the formulae (\ref{f2})
and (\ref{f1}) to this  single
solution of the Knizhnik--Zamolodchikov equation (\ref{factorizek1}) one obtains:
\begin{equation}
F_2^{(1)}=\frac{2z-1}{4\epsilon\gamma z(1-z)}, \quad F_1^{(1)}=\frac{z-2}{4\epsilon\gamma (1-z)}
\end{equation}
Analysis  of the full set of solutions of (\ref{factorizek1}) reveals
that it is only these blocks which enter the physical correlator - a fact
which is intimately related to the factorization 
of the Knizhhnik--Zamolodchikov equation (\ref{factorizek1}). In the
$su(2)$ WZNW model the factorization properties of the differential
operators are related to a finite closure of the underlying operator
algebra \cite{Christe:finite}. 

\subsubsection{Factorization at Level $k=-2$}

At level $k=-2$ ($x=4$) the differential equation (\ref{f3}) takes the
form
\begin{equation}
\begin{split}
& [4z(1-z)]^3 F_3^{'''}+9[4z(1-z)]^2(1-2z)F_3^{''} + \\ 
& \hspace{3cm} 4z(1-z)[-5+8z-48z(1-z)]F_3^{'} + \\
& \hspace{7cm} [-5+2z+8z(1-z)]F_3=0
\end{split} 
\end{equation}
This equation may be factorized to read
\begin{equation}
\label{factorizedminustwo}
{\mathcal D}^{(1)}{\mathcal D}^{(2)}F_3=0
\end{equation}
where this time the differential operators ${\mathcal D}^{(1)}$ and ${\mathcal D}^{(2)}$  take the form
\begin{eqnarray}
{\mathcal D}^{(1)} & = & 4z(1-z)\frac{d}{dz}-(5-6z) \\
{\mathcal D}^{(2)} & = & [4z(1-z)]^2 \frac{d^2}{dz^2}+8z(1-z)(3-4z)\frac{d}{dz}+(1-4z)
\end{eqnarray}
In the light of the factorization which occured at $k=1$ one might
expect that the subset of solutions obtained from ${\mathcal
D}^{(2)}F_3=0$, which are closed under analytic continuation on the
Riemann sphere, might serve as a reduced basis on which to perform the
conformal bootstrap. This indeed turns out to
be the case. It is easily seen that the equation ${\mathcal
D}^{(2)}F_3=0$, and therefore the full Knizhnik--Zamolodchikov
equation (\ref{factorizedminustwo}) admits the two
solutions\footnote{Upon the change of variables
$F_3=[z(1-z)]^{-1/4}H(z)$, the equation ${\mathcal D}^{(2)}F_3=0$
reduces to the canonical form of the hypergeometric equation 
$z(1-z)H^{''}+[c-(a+b+1)z]H^{'}-abH=0$ with $a=-1/2$, $b=1/2$,
$c=1$. This has solutions $H^{(1)}    =    \
_{2}F_{1}\left[-\tfrac{1}{2},\tfrac{1}{2};1;z\right]$ and $H^{(2)} =   (1-z)\
_{2}F_{1}\left[\tfrac{1}{2},\tfrac{3}{2};2;1-z\right]$. Using the well
known fact that $\
_{2}F_{1}\left[-\tfrac{1}{2},\tfrac{1}{2};1;z\right] =\tfrac{2}{\pi} E(z)$
together with it's derivative, the result follows.}
\begin{equation}
\label{ellipticsolutions}
F_{3}^{(1)}(z) = \frac{E(z)}{[z(1-z)]^{1/4}} \hspace{1cm} F_{3}^{(2)}(z) = \frac{E(1-z)-K(1-z)}{[z(1-z)]^{1/4}}
\end{equation}
where $K(z)$ is the complete elliptic integral of the first
kind and $E(z)$ is the complete elliptic integral of the second
kind:\footnote{Note that many texts on the theory of elliptic integrals
denote the parameter $z$ by $k^2$ - the so-called modulus. This is
purely a mater of convention.}
\begin{equation}
E(z)=\int_0^1\frac{\sqrt{1-zx^2}}{\sqrt{1-x^2}}dx\hspace{1cm} K(z)=\int_0^1\frac{1}{\sqrt{(1-x^2)(1-zx^2)}}\, dx
\end{equation}
The representation (\ref{ellipticsolutions}) is particularly useful owing to the very
simple manner in which the elliptic integrals behave under
differentiation with respect to the parameter $z$:
\begin{equation}
\frac{dE(z)}{dz}=\frac{E(z)-K(z)}{2z} \hspace{1cm} \frac{dK(z)}{dz}=\frac{E(z)-(1-z)K(z)}{2z(1-z)}
\end{equation}
Applying (\ref{f2}) and (\ref{f1}) to these solutions yields:
\begin{alignat}{2}
F_{2}^{(1)} & =  -\frac{K(z)}{4\epsilon\gamma [z(1-z)]^{1/4}} 
& \qquad 
F_{1}^{(1)} & =  \frac{z K(z)}{4\epsilon \gamma [z(1-z)]^{1/4}} \label{blocks1}
\\
F_{2}^{(2)} & = \frac{K(1-z)}{4\epsilon \gamma [z(1-z)]^{1/4}} 
& \qquad 
F_{1}^{(2)} & = -\frac{z K(1-z)}{4 \epsilon \gamma [z(1-z)]^{1/4}}\label{blocks2}
\end{alignat}
 As we shall subsequently demonstrate, one may satisfy the demands of
single-valuedness and crossing symmetry on the subspace of functions
(\ref{ellipticsolutions}), (\ref{blocks1}) and (\ref{blocks2}). Once
again this is intimately connected with the factorization of the
Knizhnik--Zamolodchikov equation (\ref{factorizedminustwo}).
Having found closed form expressions for the chiral blocks, one must
now construct the physical correlation functions to be single-valued
on the whole Riemann sphere. It is enough to ensure this property at
the two singular points $z=0$ and $z=1$. The blocks $F_{i}^{(1)}(z)$ are regular
at $z=0$ and logarithmic at $z=1$, whilst the blocks $F_{i}^{(2)}(z)$ are
logarithmic at $z=0$ and regular at $z=1$. It is straightforward to see
that in this subspace of functions one can only have\footnote{A detailed proof of this statement in terms of monodromy matrices is given in appendix \ref{app:monod}.}
\begin{equation}
F_{ij}(z, \bar z) = X_{12}\left[F_i^{(1)}(z)F_j^{(2)}({\bar z})+F_i^{(2)}(z)F_j^{(1)}({\bar z})\right]
\end{equation}
 One may gather the explicit expressions for the $F_{ij}$ at level
 $k=-2$ in the Hermitian matrix:
\begin{align}
\label{fmatrix}
F_{ij}& =\begin{pmatrix}|z|^2F_{22} & -zF_{22} & -zF_{23} \\ -\bar z
F_{22} & F_{22} & F_{23}\\ -{\Bar z}{\Bar F}_{23} & {\Bar F}_{23} & F_{33}
\end{pmatrix}
\intertext{in which we have singled out the elements}
F_{22} & = -\Lambda\left[K\Kbartilde+\Kbar\Ktilde\right] \label{theta}\\
F_{23} & = 4\epsilon\gamma\Lambda\left[K\Kbartilde+\Ktilde\Ebar-K\Ebartilde\right]\label{omega}\\
F_{33} & = (4\epsilon\gamma)^2\Lambda\left[E\Ebartilde-E\Kbartilde+\Ebar\Etilde-\Ebar\Ktilde\right]\label{sigma}\\
\intertext{where}
\Lambda & = \frac{X_{12}}{(4\epsilon\gamma)^2|z(1-z)|^{1/2}}\label{lambda}
\end{align}
and where we have adopted the notation that  ${\Tilde f}(z)\equiv f(1-z)$, ${\bar f}(z)\equiv f({\Bar
z})$ and ${\Tilde {\Bar f}}(z)\equiv f(1-{\Bar z})$ for an arbitrary
function $f$. One may also demonstrate that this combination is consistent
with the crossing symmetry constraints on the correlator:
\begin{equation}
F^{\boldsymbol{\alpha}, \boldsymbol{\bar \alpha}}(z,\bar z)  = 
{\Tilde {\mathcal P}}{\Tilde {\Bar {\mathcal P}}}F^{\boldsymbol{\tilde\alpha}, \boldsymbol{\tilde{\bar\alpha}}}(1-z,1-\bar z),\qquad
F^{\boldsymbol{\alpha}, \boldsymbol{\bar \alpha}}(z,\bar z)  = 
z^{-2\Delta}{\bar z}^{-2\Delta}{\Hat {\mathcal P}}{\Hat {\Bar {\mathcal P}}}    F^{\boldsymbol{\hat\alpha}, \boldsymbol{\hat{\bar\alpha}}}(1/z,1/\bar z)
\end{equation}
where $\boldsymbol{\tilde\alpha}$ denotes the permuted sequence of
indices $\alpha_1,\alpha_3,\alpha_2,\alpha_4$,
$\boldsymbol{\hat\alpha}$ denotes the sequence
$\alpha_1,\alpha_4,\alpha_3,\alpha_2$, and ${\mathcal P}$ denotes the parity of the
permutation:
\begin{equation}
{\Tilde {\mathcal P}}  = (-1)^{\varepsilon_{\alpha_2}\varepsilon_{\alpha_3}},\qquad
{\Hat {\mathcal P}}  =  (-1)^{\varepsilon_{\alpha_2}(\varepsilon_{\alpha_3}+\varepsilon_{\alpha_4})+\varepsilon_{\alpha_3}\varepsilon_{\alpha_4}}
\end{equation}
The proof of this requires the use of the analytic continuation
formulae of the elliptic integrals and is presented in appendix
\ref{app:cross}.

\subsection{Monodromy Invariance}
\label{app:monod}
A monodromy transformation of a function of z consists in letting z
circulate around some other point (typically a singular
point). We define
\begin{eqnarray}
{\mathcal C}_0\,F(z,\bar z) & = & \lim_{t\rightarrow 1^-}F(ze^{2i\pi
t},\bar ze^{-2i\pi t}) \\
{\mathcal C}_1\,F(z,\bar z) & = & \lim_{t\rightarrow 1^-}F(1+(z-1)e^{2i\pi t},1+(\bar z-1)e^{-2i\pi t})
\end{eqnarray}
Using the standard analytic continuation formulae for the
hypergeometric series, it is easily seen that the elliptic integrals have
the following nontrivial monodromy properties:
\begin{eqnarray}
{\mathcal C}_0 K(1-z) & = & K(1-z)-2iK(z)   \\
{\mathcal C}_0 E(1-z) & = & E(1-z)+2i[E(z)-K(z)]  \\
{\mathcal C}_1 K(z) & = & K(z)-2iK(1-z)  \\
{\mathcal C}_1 E(z) & = & E(z)+2i[E(1-z)-K(1-z)]    
\end{eqnarray}
together with the trivial transformations ${\mathcal C}_0 K(z)  = 
K(z)$, ${\mathcal C}_0 E(z)  = E(z)$, ${\mathcal C}_1 K(1-z)  = 
K(1-z)$, ${\mathcal C}_1 E(1-z)  = 
E(1-z)$. Using these results it is straightforward to see that
\begin{eqnarray}
{\mathcal C}_0\, F_i^{(a)}(z) & = &  (g_0)_{ab}\,F_i^{(b)}(z) , \quad i=1,2,3\\
{\mathcal C}_1\, F_i^{(a)}(z) & = &  (g_1)_{ab}\,F_i^{(b)}(z) , \quad i=1,2,3
\end{eqnarray}
where (on this reduced subspace) the matrices $g_0$ and $g_1$ are given by
\begin{equation}
g_0=\begin{pmatrix} -i & 0 \\ 2 &
    -i  \\ 
	\end{pmatrix}, \quad g_1=\begin{pmatrix} -i & 2 \\ 0 &
    -i \\
	\end{pmatrix}.
\end{equation}
Under the monodromy transformation ${\mathcal C}_0$, the combination
\begin{equation}
F_{ij}(z, \bar z)=\sum_{a,b=1}^2X_{ab}F_i^{(a)}(z)F_j^{(b)}(\bar z)
\end{equation}
transforms in the following manner
\begin{align}
{\mathcal C}_0F_{ij}(z, \bar z) & =  F_i^{(1)}(z)F_j^{(1)}({\bar
z})\left[X_{11}-2i(X_{12}-X_{21})+4X_{22}\right]+F_i^{(2)}(z)F_j^{(2)}({\bar
z})\left[X_{22}\right]\notag \\
& +  F_i^{(1)}(z)F_j^{(2)}({\bar z})\left[X_{12}+2iX_{22}\right]+ F_i^{(2)}(z)F_j^{(1)}({\bar
z})\left[X_{21}-2iX_{22}\right]\notag
\end{align}
Invariance under the monodromy transformation ${\mathcal C}_0$ thus
requires $X_{12}=X_{21}$, and
$X_{22}=0$. That is to say
\begin{equation}
F_{ij}(z, \bar z)=X_{11}F_i^{(1)}(z)F_j^{(1)}({\bar z})+X_{12}\left[F_i^{(1)}(z)F_j^{(2)}({\bar z})+F_i^{(2)}(z)F_j^{(1)}({\bar z})\right]\notag
\end{equation}
Under the monodromy transformation ${\mathcal C}_1$ this simplified
function transforms as
\begin{align}
{\mathcal C}_1F_{ij}(z, \bar z) & =  F_i^{(1)}(z)F_j^{(1)}({\bar
z})\left[X_{11}\right]+F_i^{(1)}(z)F_j^{(2)}({\bar
z})\left[X_{12}-2iX_{11}\right] \notag \\
& + F_i^{(2)}(z)F_j^{(1)}({\bar
z})\left[X_{12}+2iX_{11}\right]+F_i^{(2)}(z)F_j^{(2)}({\bar
z})\left[4X_{11}\right]\notag
\end{align}
Invariance under the monodromy transformation ${\mathcal C}_1$
therefore imposes the additional constraint $X_{11}=0$. Hence
monodromy invariance restricts $F_{ij}(z,\bar z)$ to have the form
\begin{equation}
F_{ij}(z, \bar z)=X_{12}\left[F_i^{(1)}(z)F_j^{(2)}({\bar z})+F_i^{(2)}(z)F_j^{(1)}({\bar z})\right].
\end{equation}
as stated in the text. 
\subsection{Crossing Symmetry}
\label{app:cross}
\subsubsection{Invariance under $z\rightarrow 1-z$}
Crossing symmetry requires that
\begin{equation}
F^{\boldsymbol{\alpha}, \boldsymbol{\bar \alpha}}(z,\bar z)  = 
{\tilde {\mathcal P}}{\Tilde {\Bar {\mathcal P}}}F^{\boldsymbol{\tilde\alpha}, \boldsymbol{\tilde{\bar\alpha}}}(1-z,1-\bar z)
\end{equation}
where $\boldsymbol{\alpha}$ denotes the sequence of indices $\alpha_1,\alpha_2,\alpha_3,\alpha_4$, $\boldsymbol{\Tilde\alpha}$ denotes the permuted sequence of
indices $\alpha_1,\alpha_3,\alpha_2,\alpha_4$, ${\Tilde
{\mathcal P}}=(-1)^{\varepsilon_{\alpha_2}\varepsilon_{\alpha_3}}$ denotes the parity
of the interchange in the holomorphic sector ($\varepsilon_{\alpha}$ is
$0$ for bosons and $1$ for fermions.) Introducing the following tensor \cite{Maassarani:non}
\begin{equation}
\label{jtensor}
J_{i}^{\boldsymbol{\alpha}}={\Tilde {\mathcal P}} I_{i}^{\Tilde{\boldsymbol{\alpha}}}
\end{equation}
the crossing symmetry constraint may be written
\begin{equation}
\label{crossconst}
\sum_{i,j=1}^3 I_{i}^{\boldsymbol \alpha} {\bar
I}_{j}^{\boldsymbol{\Bar\alpha}} F_{ij}(z,\bar z)=\sum_{i,j=1}^3
J_{i}^{\boldsymbol \alpha} {\bar J}_{j}^{\boldsymbol{ \bar\alpha }} F_{ij}(1-z,1-\bar z)
\end{equation}
The tensor $J$ admits the following decomposition \cite{Maassarani:non}
\begin{equation}
\label{jitensorcorr}
J_i^{\boldsymbol \alpha}=C_{1}^{ij}I_{j}^{\boldsymbol \alpha}\quad C_1=\begin{pmatrix} -1 & 0 & 0 \\ -1& 1 & -4\epsilon\gamma \\ 0 & 0 & -1 \\ \end{pmatrix}
\end{equation}
Substituting this decomposition into equation (\ref{crossconst}) and equating the coefficients of $I_{i}{\bar I}_{j}$ on both sides, one finds the following nine identities which must be satisfied by the $F_{ij}(z,\bar z)$ if this crossing symmetry is to be satisfied. Denoting $F_{ij}(z,\bar z)$ by $F_{ij}$, and $F_{ij}(1-z,1-\bar z)$ by ${\tilde F}_{ij}$ these are as  follows:
\begin{subequations}
\begin{eqnarray}
F_{11} & = & {\tilde F}_{11}+{\tilde F}_{12}+{\tilde F}_{21}+{\tilde F}_{22} \\
F_{12} & = & -{\tilde F}_{12}-{\tilde F}_{22} \\
F_{13} & = & 4\epsilon\gamma ({\tilde F}_{12}+{\tilde F}_{22})+{\tilde F}_{13}+{\tilde F}_{23}  \\
F_{21} & = & -{\tilde F}_{21}-{\tilde F}_{22} \\
F_{22} & = & {\tilde F}_{22} \\
F_{23} & = & -4\epsilon\gamma{\tilde F}_{22}-{\tilde F}_{23} \\
F_{31} & = & 4\epsilon\gamma ({\tilde F}_{21}+{\tilde F}_{22})+{\tilde F}_{31}+{\tilde F}_{32} \\
F_{32} & = & -4\epsilon\gamma{\tilde F}_{22}-{\tilde F}_{32}\\
F_{33} & = & 4\epsilon\gamma(4\epsilon\gamma{\tilde F}_{22}+{\tilde
F}_{23}+{\tilde F}_{32})+{\tilde F}_{33} \label{ztooneminus}
\end{eqnarray}
\end{subequations}
It is straightforward to show that these relations are indeed satisfied. 
\subsubsection{Invariance under $z\rightarrow 1/z$}
Crossing symmetry requires that
\begin{equation}
F^{\boldsymbol{\alpha}, \boldsymbol{\bar \alpha}}(z,\bar z)  =
z^{-2\Delta}{\bar z}^{-2\Delta}
{\Hat{\mathcal P}}{\Hat {\Bar {\mathcal P}}}F^{\boldsymbol{\Hat\alpha}, \boldsymbol{\Hat{\bar\alpha}}}(1/z,1/\bar z)
\end{equation}
where $\boldsymbol{\alpha}$ denotes the sequence of indices $\alpha_1,\alpha_2,\alpha_3,\alpha_4$, $\boldsymbol{\Hat\alpha}$ denotes the permuted sequence of
indices $\alpha_1,\alpha_4,\alpha_3,\alpha_2$, ${\Hat
{\mathcal P}}=(-1)^{\varepsilon_{\alpha_2}(\varepsilon_{\alpha_3}+\varepsilon_{\alpha_4})+\varepsilon_{\alpha_3}\varepsilon_{\alpha_4}}$ denotes the parity
of the interchange in the holomorphic sector ($\varepsilon_{\alpha}$ is
$0$ for bosons and $1$ for fermions.) Introducing the following tensor \cite{Maassarani:non}
\begin{equation}
K_{i}^{\boldsymbol{\alpha}}={\Hat {\mathcal P}} I_{i}^{\Hat{\boldsymbol{\alpha}}}
\end{equation}
the crossing symmetry constraint may be written
\begin{equation}
\label{crosssymm}
\sum_{i,j=1}^3 I_{i}^{\boldsymbol \alpha} {\bar I}_{j}^{\boldsymbol{\bar\alpha}} F_{ij}(z,\bar z)=z^{-1/x}{\bar z}^{-1/x}\sum_{i,j=1}^3 K_{i}^{\boldsymbol \alpha} {\bar K}_{j}^{\boldsymbol{ \bar\alpha}} F_{ij}(1/z,1/\bar z)
\end{equation}
 The tensor $K$ admits the following decomposition \cite{Maassarani:non}
\begin{equation}
K_i^{\boldsymbol \alpha}=C_{2}^{ij}I_{j}^{\boldsymbol \alpha}\quad C_2=\begin{pmatrix} 0 & 1 & -4\epsilon\gamma \\ 1& 0 & -4\epsilon\gamma \\ 0 & 0 & -1 \\ \end{pmatrix}
\end{equation}
Substituting this decomposition into equation (\ref{crosssymm}) and
equating the coefficients of $I_{i}{\bar I}_{j}$ on both sides, one
finds the following nine identities which must be satisfied by the
$F_{ij}(z,\bar z)$ if this crossing symmetry is to be
satisfied. Denoting $F_{ij}(z,\bar z)$ by $F_{ij}$, and
$F_{ij}(1/z,1/\bar z)$ by ${\hat F}_{ij}$ these are
as  follows:
\begin{subequations}
\begin{eqnarray}
F_{11} & = & |z|^{-2/x}{\hat F}_{22} \label{f11}\\
F_{12} & = & |z|^{-2/x}{\hat F}_{21}  \label{f12}\\
F_{13} & = & |z|^{-2/x}[ -4\epsilon\gamma({\hat F}_{21}+{\hat F}_{22})-{\hat F}_{23}]\label{f13}\\
F_{21} & = &  |z|^{-2/x}{\hat F}_{12} \label{f21}\\
F_{22} & = &  |z|^{-2/x}{\hat F}_{11} \label{f22}\\
F_{23} & = &  |z|^{-2/x}[-4\epsilon\gamma({\hat F}_{11}+{\hat F}_{12})-{\hat F}_{13}] \label{f23}\\
F_{31} & = &  |z|^{-2/x}[ -4\epsilon\gamma({\hat F}_{12}+{\hat F}_{22})-{\hat F}_{32}] \label{f31}\\
F_{32} & = &  |z|^{-2/x}[-4\epsilon\gamma({\hat F}_{11}+{\hat F}_{21})-{\hat F}_{31}] \label{f32}\\
F_{33} & = &  |z|^{-2/x}[16\epsilon^2\gamma^2({\hat F}_{11}+{\hat
F}_{12}+{\hat F}_{21}+{\hat F}_{22})+  \notag \\
& & \hspace{1.5cm} 4\epsilon\gamma({\hat F}_{13}+{\hat F}_{23}+{\hat F}_{31}+{\hat F}_{32})+{\hat F}_{33}] \label{f33}
\end{eqnarray}
\end{subequations}
In order to demonstrate that these identities are satisfied we shall
make use of the following rather simple transformation laws of the
elliptic integrals under $z\rightarrow 1/z$:\footnote{We note that replacing
$i$ by $-i$ in equations (\ref{kover}) and (\ref{eover}) changes the
domain of validity of the transformation from ${\mathfrak Im} z <0$
to ${\mathfrak Im} z > 0$. Since $E$ and $K$ appear together in the
$F_{ij}$ it is essential that their transformations be defined in the
same region of the complex plane.}
\begin{subequations}
\begin{eqnarray}
K(1/z) & = & z^{+1/2}\left[K(z)+iK(1-z)\right], \quad {\mathfrak
Im} z <0 \label{kover}\\
E(1/z) & = & z^{-1/2}\left[D(z)-iD(1-z)\right], \quad {\mathfrak
Im} z <0 \label{eover}\\
K(1-1/z) & = & z^{+1/2}K(1-z) \label{kminusover} \\
E(1-1/z) & = & z^{-1/2}E(1-z) \label{eminusover}
\end{eqnarray}
\end{subequations}
where
\begin{equation}
D(z)=E(z)-(1-z)K(z).
\end{equation}
These are easily obtained using the standard analytic continuation
formulae for ordinary hypergeometric functions. Recalling the form of
$F_{22}$ appearing in equation (\ref{theta}) namely
\begin{equation}
F_{22}=-\Lambda \left[K(1-z)K({\Bar z})+K(z)K(1-{\Bar z})\right]
\end{equation}
it is straightforward to see how it
transforms under $z\rightarrow1/z$:
\begin{equation}
\label{hatf22}
{\Hat F}_{22}  = |z|^{5/2}F_{22}
\end{equation}
Since  $F_{11}=|z|^2F_{22}$ from equation (\ref{fmatrix}) it follows that
that  constraint (\ref{f11}) is satisfied. Replacing now $z$ by $1/z$
in (\ref{f11}) one may infer the validity of (\ref{f22}). Further, since
$F_{21}=-{\bar z}F_{22}$
from equation (\ref{fmatrix}) it follows by reciprocity
that 
\begin{equation}
{\Hat F}_{21}=-{\bar z}^{-1}{\Hat
F}_{22}.
\end{equation}
Substituting for ${\Hat F}_{22}$ using (\ref{hatf22}) and using the
fact that $F_{12}=-zF_{22}$ one obtains
\begin{equation}
{\Hat F}_{21}=|z|^{1/2}F_{12}
\end{equation}
That is to say, constraint (\ref{f12}) is satisfied. By
reciprocity we see that  constraint (\ref{f21}) is also satisfied.
Recalling now the form of $F_{23}$ appearing in equation (\ref{omega})
namely
\begin{equation}
F_{23}=4\epsilon\gamma\Lambda\left[K\Kbartilde+\Ktilde\Ebar-K\Ebartilde\right]
\end{equation}
it is straightforward to see how it
transforms under $z\rightarrow1/z$:
\begin{eqnarray}
{\Hat F}_{23} & = &
4\epsilon\gamma|z|^{3/2}\Lambda\left[z^{1/2}\left(K+i\Ktilde\right){\Bar
z}^{1/2}\Kbartilde+\right.\notag \\
& & \hspace{3cm}\left.z^{1/2}\Ktilde{\Bar z}^{-1/2}\left({\Bar
D}+i{\Tilde{{\Bar D}}}\right)-z^{1/2}\left(K+i\Ktilde\right){\Bar
z}^{-1/2}\Ebartilde\right] \\
& = &
4\epsilon\gamma|z|^{1/2}\Lambda\left[|z|^2\left(K+i\Ktilde\right)\Kbartilde+z\Ktilde\left(\Ebar-(1-{\bar
z})\Kbar\right)\right. \notag \\ 
& & \hspace{5.2cm}\left. +i z\Ktilde\left(\Ebartilde-{\bar
z}\Kbartilde\right)-z\left(K+i\Ktilde\right)\Ebartilde\right]\\
& = &
4\epsilon\gamma|z|^{1/2}\Lambda\left[|z|^2\left(K\Kbartilde+\Ktilde\Kbar\right)+z\left(\Ktilde\Ebar-\Ktilde\Kbar-K\Ebartilde\right)\right]\\
& = &
4\epsilon\gamma|z|^{1/2}\Lambda\left[|z|^2\left(K\Kbartilde+\Ktilde\Kbar\right)-z\left(K\Kbartilde+\Ktilde\Kbar\right)
+z\left(K\Kbartilde+\Ktilde\Ebar-K\Ebartilde\right)\right]\\
& = &
4\epsilon\gamma|z|^{1/2}\left[-|z|^2F_{22}+zF_{22}+\frac{z}{4\epsilon\gamma}F_{23}\right]\\
& = & |z|^{1/2}\left[-4\epsilon\gamma(F_{11}+F_{12})-F_{13}\right] \label{f23hat}
\end{eqnarray}
Replacing $z$ by $1/z$ on both sides of this equation one obtains the
relation (\ref{f23}). Taking now the complex conjugate of (\ref{f23})
and using the Hermiticity property ${\Bar F}_{ij}=F_{ji}$, one obtains
(\ref{f32}). Further, rearranging (\ref{f23hat}) for $F_{13}$ and
replacing $F_{11}$ and $F_{12}$ using (\ref{f11}) and (\ref{f12})
respectively one obtains (\ref{f13}). Taking  now the complex conjugate of (\ref{f13})
and using Hermiticity, one obtains
(\ref{f31}).  We consider finally how $F_{33}$, namely
\begin{equation}
F_{33} = (4\epsilon\gamma)^2\Lambda\left[E(\Ebartilde-\Kbartilde)+ {\rm c.c.}\right]
\end{equation}
behaves under the transformation $z\rightarrow1/z$:
\begin{eqnarray}
{\Hat F}_{33} & = &
(4\epsilon\gamma)^2|z|^{3/2}\Lambda\left[z^{-1/2}(D-i{\Tilde D})({\Bar
z}^{-1/2}\Ebartilde-{\Bar z}^{1/2}\Kbartilde)+c.c.\right]\\
& =& (4\epsilon\gamma)^2|z|^{1/2}\Lambda\left[(D-i{\Tilde
D})(\Ebartilde-{\Bar z} \Kbartilde)+c.c.\right]\\
& = &
(4\epsilon\gamma)^2|z|^{1/2}\Lambda\left[E\Ebartilde-(1-z)K\Ebartilde-z\Ebar\Ktilde+{\Bar
z}(1-z)K\Kbartilde+c.c.\right]\\
& = &
(4\epsilon\gamma)^2|z|^{1/2}\Lambda\left[E\Ebartilde-\Ebar\Ktilde+(1-z)(\Ebar\Ktilde-K\Ebartilde+K\Kbartilde)-|1-z|^2K\Kbartilde+c.c.\right]
\\
& = &
|z|^{1/2}\left[(4\epsilon\gamma)^2|1-z|^2F_{22}+4\epsilon\gamma\left[(1-z)F_{23}+c.c.\right]+F_{33}\right]
\\
& = & |z|^{1/2}\left[(4\epsilon\gamma)^2(F_{11}+F_{12}+F_{21}+F_{22})+4\epsilon\gamma(F_{13}+F_{23}+F_{31}+F_{32})+F_{33}\right]
\end{eqnarray}
which may be seen to be the relation (\ref{f33}) with $z$ replaced by
$1/z$. This completes our proof of crossing symmetry in the reduced
subspace.

\end{document}